\documentclass[aps,prd,showpacs,floatfix,preprintnumbers,groupedaddress]{revtex4}
\usepackage{graphicx}
\usepackage{psfrag}
\newcommand{\sla}[1]{/\!\!\!#1}
\def\lsim{\raise0.3ex\hbox{$\;<$\kern-0.75em\raise-1.1ex\hbox{$\sim\;$}}}
\def\gsim{\raise0.3ex\hbox{$\;>$\kern-0.75em\raise-1.1ex\hbox{$\sim\;$}}}

\def\hbar{\hspace{0pt}\raisebox{1pt}{$-$} \hspace{-7pt} h}

\newcommand{\be}{\begin{equation}}
\newcommand{\ee}{\end{equation}}
\newcommand{\bd}{\begin{displaymath}}
\newcommand{\ed}{\end{displaymath}}
\newcommand{\bea}{\begin{eqnarray}}
\newcommand{\eea}{\end{eqnarray}}

\newcommand {\ignore}[1]{}

\def\10{SO(10)}
\def\321{SU(3) $\otimes$ SU(2) $\otimes$ U(1) }

\newcommand{\AddrAHEP}{%
  AHEP Group, Institut de F\'{\i}sica Corpuscular --
  C.S.I.C./Universitat de Val{\`e}ncia \\
  Edificio Institutos de Paterna, Apt 22085, E--46071 Valencia, Spain}

\newcommand{\AddrWur}{%
Institut f\"ur Theoretische Physik und Astronomie
Universit\"at W\"urzburg\\
Am Hubland
97074 Wuerzburg}
\begin{document}
\preprint{IFIC/07-50}
\title{Probing bilinear R--parity violating supergravity at the LHC}
\date{\today}

\author{F.\ de Campos}
\email{camposc@feg.unesp.br}
\affiliation{Departamento de F\'{\i}sica e Qu\'{\i}mica,
             Universidade Estadual Paulista, \\
             Av. Dr. Ariberto Pereira da Cunha, 333, 12516-410, 
             Guaratinguet\'a -- SP,  Brazil }

\author{O.\ J.\ P.\ \'Eboli}
\email{eboli@fma.if.usp.br}
\affiliation{Instituto de F\'{\i}sica,
             Universidade de S\~ao Paulo, \\
             C.P. 66318, 05315-970, S\~ao Paulo -- SP, Brazil.}

\author{M.\ B.\ Magro} \email{magro@fma.if.usp.br}
\affiliation{Faculdade de Engenharia,
             Centro Universit\'ario Funda\c{c}\~ao Santo Andr\'e, \\
             Av. Pr\'{\i}ncipe de Gales, 821, 09060-650, 
             Santo Andr\'e -- SP, Brazil.}

\author{W.\ Porod} \email{porod@ific.uv.es}
\affiliation{\AddrWur}

\author{D.\ Restrepo} \email{restrepo@uv.es}
\affiliation{Instituto de F\'{\i}sica, Universidad de Antioquia \\
             A.A. 1226, Medellin, Colombia}

\author{M.~Hirsch} \email{mahirsch@ific.uv.es} \affiliation{\AddrAHEP}

\author{J.~W.~F.~Valle} \email{valle@ific.uv.es} \affiliation{\AddrAHEP}


\begin{abstract}
\vspace*{1cm}

We study the collider phenomenology of bilinear R--parity violating
supergravity, the simplest effective model for supersymmetric neutrino
masses accounting for the current neutrino oscillation data. At the
CERN Large Hadron Collider the center--of--mass energy will be high
enough to probe directly these models through the search for the
superpartners of the Standard Model (SM) particles. We analyze the impact
of R--parity violation on the canonical supersymmetry searches
-- that is, we examine how the decay of the lightest supersymmetric
particle (LSP) via bilinear R--parity violating interactions degrades
the average expected missing momentum of the reactions and show how
this diminishes the reach in the \emph{usual} channels for
supersymmetry searches. However, the R--parity violating interactions
lead to an enhancement of the final states containing isolated
same-sign di-leptons and trileptons, compensating the reach loss in
the fully inclusive channel. We show how the searches for
\emph{displaced vertices} associated to LSP decay substantially
increase the coverage in supergravity parameter space, giving the
corresponding reaches for two reference luminosities of 10 and 100
fb$^{-1}$ and compare with those of the R--parity conserving minimal
supergravity model.

\end{abstract}

\maketitle


\centerline{\today}

\section{Introduction}

Weak scale supersymmetry (SUSY) represents the most popular approach
towards a solution of the hierarchy problem, with the added virtue of
being perturbative and accounting for the unification of the gauge
coupling constants and the possibility of eventually including
gravity~\cite{Martin:1997ns}.
Little is currently known about the detailed mechanism of how SUSY is
realized and of how it is broken. Ultimately only experiment can
settle this issue.
The most popular ansatz \textsl{assumes} that R--parity is conserved
and that SUSY breaking in a hidden sector is communicated to the
observable sector by flavour-blind gravitational interactions.
Neither of these assumptions is mandatory, in fact they are both {\it
  ad hoc}. Many different mechanisms of supersymmetry breaking can be
envisaged~\cite{Brignole:1994dj,Randall:1998uk,giudice:1998xp} and,
similarly, the breaking of R--parity~\cite{Fayet:1976cr} can take
place in a variety of
ways~\cite{hall:1984id,ross:1985yg,ellis:1985gi}. 

Here we will adopt only the first part of the conventional ansatz,
namely assume transmission of supersymmetry  breaking via
gravity but relax the R--parity conservation assumption.
A specially interesting possibility is that R--parity breaks
spontaneously, as a result of the minimization of the scalar Higgs
potential~\cite{masiero:1990uj}, very much in the same way as the
electroweak symmetry itself breaks.
Spontaneous R-parity breaking models are characterized by two types of
sneutrino vacuum expectation values (vev)~\footnote{Left-handed
  sneutrino vevs were considered in the pre--LEP
  days~\cite{ross:1985yg}.  After the LEP measurements of the
  invisible Z-width it became clear that the existence of gauge
  singlet ``right-handed'' sneutrino vevs is necessary.}, those of
right and left-handed sneutrinos, singlets and doublets under \321
respectively, obeying the ``vev-seesaw'' relation $v_L v_R \sim h_\nu
m_W^2$ where $h_\nu$ is the small Yukawa coupling that governs the
strength of R--parity violation~\cite{masiero:1990uj}.
Below the scale $v_R$ the effective breaking of R--parity is explicit
through bilinear superpotential terms and the corresponding soft
supersymmetry breaking terms.
For generality and simplicity here we replace the right-handed
sneutrino $v_R$ vev by the effective bilinear coupling.

We refer to such simple effective bilinear R--parity breaking (BRpV)
picture as BRpV--mSUGRA for short~\cite{Diaz:1997xc}~\cite{chun:1998gp}. 
Here we start directly from the BRpV--mSUGRA model, which has a strong
phenomenological motivation, besides simplicity, namely that it
provides the most economical effective model for supersymmetric
neutrino masses, that successfully accounts for the observed pattern
of neutrino masses and
mixings~\cite{Diaz:2003as,Hirsch:2000ef}. Moreover, it has been shown
that, despite the smallness of R--parity violation indicated by the
neutrino oscillation data, its effects can be tested and the model
falsified at collider experiments~\cite{Hirsch:2004he}. Here we focus
on the implications of the model for the first stages of the CERN Large
Hadron Collider (LHC).

There have been extensive studies of searches for supersymmetric
particles but mainly in the context of models with conserved R
parity~\cite{LHCTDR,LC,Allanach:2002nj,AguilarSaavedra:2005pw}.
So far no direct experimental evidence for SUSY in high energy
experiments has been found.
With the coming into operation of the LHC searches for R--parity
breaking
signatures~\cite{Magro:2003zb,deCampos:2005ri,allanach:1999bf,Barbier:2004ez,
  rplist} acquire stronger motivation not only by their intrinsic
interest, but also because they open the exciting possibility of
probing the neutrino oscillation angles at collider
experiments~\cite{Hirsch:2003fe,hirsch:2002ys,Restrepo:2001me,Porod:2000hv,romao:1999up,mukhopadhyaya:1998xj,Choi:1999tq}.


Previously we have studied the Tevatron potential to search for
BRpV--mSUGRA in the multilepton channel~\cite{Magro:2003zb} as well as in
the form of displaced vertices~\cite{deCampos:2005ri}.
In this work we provide a quantitative analysis of the
phenomenological implications of bilinear R--parity breaking
supergravity for the LHC, extending our studies to the new energy
range accessible at the LHC and including all constraints from neutrino
oscillation physics.
Since the effective strength of R--parity violation must be small in
order to fit current neutrino oscillation data, our study also
constitutes a robustness check of the supergravity parameter reach
estimates against the presence of ``perturbative'' BRpV terms.
Supersymmetric particle spectra and production cross sections are
expected to be the same as the conventional ones, thanks to the
smallness of R--parity violation required by neutrino oscillation
data.  Similarly, processes such as $b \to s\gamma$ and g-2 are
essentially the same in BRpV--mSUGRA as in mSUGRA and hence the resulting
constraints for the latter still hold.

The basic difference between BRpV--mSUGRA and the conventional R-parity
conserving mSUGRA scenario is that the lightest supersymmetric particle
is no longer stable and it decays, typically inside the detector,
consequently modifying the SUSY phenomenology at colliders. This has
two important implications:
\begin{itemize}

\item the neutralino is no longer a viable dark matter candidate; one
  needs to implement either an axion-like,
  gravitino~\cite{Borgani:1996ag,Takayama:2000uz,Hirsch:2005ag,Axion} or
  possibly majoron dark matter
  scheme~\cite{Berezinsky:1993fm,Lattanzi:2007ux}. Clearly no ``SUSY
  dark matter'' constraint should be applied in these models.

\item any SUSY particle can be the lightest one, hence it can be
  electrically charged.  For the case of interest assumed here it
  means that the parameters that lead to stau-LSP are perfectly viable
  for the searches, and are indeed included in our analysis.

\end{itemize}

As far as searches go, the LSP decay implies that the missing
transverse momentum of the events gets diminished and,
correspondingly, an increase of the multiplicities for jets and/or
charged leptons in the final state is expected.  Most remarkably,
within a large range of BRpV--mSUGRA parameters, the LSP can live long
enough to give rise to displaced vertices, opening a new window for
its search and study.
%
%
An interesting phenomenological feature of BRpV not present in generic
RPV schemes is that it naturally leads to displaced
vertices~\footnote{Notice that displaced vertices may also arise in
  other scenarios of R--parity violation provided the trilinear
  couplings are sufficiently small; see for
  instance~\cite{Kaplan:2007ap}.}.

In this work we have analyzed in detail the impact of the LSP decay on
the potential of the LHC to unravel the existence of supersymmetry in
the simplest framework where its breaking is gravitational.
To achieve this we start from the same conventional R--parity
conserving search channels. We demonstrate that the LHC reach is
reduced in the jets and missing transverse momentum channel with
respect to the R--parity conserving scenario.
Furthermore, we also find that the final state topologies containing
isolated charged same-sign di-leptons and trileptons are enhanced,
partially compensating the loss of missing momentum brought about by
the LSP decay in the all inclusive channel.
An especially relevant fact is that the LSP decay can also lead to
displaced vertices. We have also investigated the capabilities of the
LHC experiments, ATLAS and CMS, to unravel the existence of SUSY with
R--parity violation through the search for displaced vertices with a
high invariant mass associated to them. In fact, we find that at 100
fb$^{-1}$ the reach in the displaced vertices is typically larger than
the one expected in the R--parity conserving scenario, especially at
large $m_0$ values.

\section{The BRpV--mSUGRA model}
\label{sec:model}

As already mentioned, here we focus on the phenomenology of the
effective theory described by the bilinear R-parity breaking
supergravity model in which the MSSM superpotential is supplemented by
the following terms~\cite{Diaz:1997xc}
\begin{equation}
W_{\text{BRpV}} = W_{\text{MSSM}}  + \varepsilon_{ab}
\epsilon_i \widehat L_i^a\widehat H_u^b \; ,
\end{equation}
with three extra parameters ($\epsilon_i$), one for each fermion
generation. In addition to these bilinear terms, we must also include
new soft supersymmetry breaking terms ($B_i$).
\begin{equation}
V_{\text{soft}} = V_{\text{MSSM}} 
-\varepsilon_{ab}\left(
B_i\epsilon_i\widetilde L_i^aH_u^b\right) \; .
\end{equation}
Taken together, the new terms in the BRpV Lagrangian (the three
bilinear $\epsilon_i$ and the $B_i$) lead to the explicit violation of
lepton number as well as R--parity. These terms also induce vevs $v_i
\equiv v_{Li}$, $i=1,2,3$ for the three left-type sneutrinos.  Note,
that in general there is no basis where both sets of bilinear R-parity
breaking terms can be eliminated at the same time.  In the basis where
the bilinear terms in the superpotential are rotated away one would
obtain a model where the trilinear terms are approximately given by
$\lambda_{ijk} = (\epsilon_i/\mu) Y_{E,jk}$ and $\lambda_{ijk}' =
(\epsilon_i/\mu) Y_{D,jk}$ \cite{Aristizabal Sierra:2004cy}.

To further specify the model we assume a minimal supergravity
framework with universal soft supersymmetry breaking terms at the
unification scale.  This model contains eleven free
parameters, namely
\begin{equation}
m_0\,,\, m_{1/2}\,,\, \tan\beta\,,\, {\mathrm{sign}}(\mu)\,,\, 
A_0 \,,\, 
\epsilon_i \: {\mathrm{, and}}\,\, \Lambda_i\,,
\end{equation}
where $m_{1/2}$ and $m_0$ are the common gaugino mass and scalar soft
SUSY breaking masses at the unification scale, $A_0$ is the common
trilinear term, and $\tan\beta$ is the ratio between the Higgs field
vev's. For convenience, we trade the soft parameters $B_i$ by the
``alignment'' parameters $\Lambda_i=\epsilon_iv_d+\mu v_i$ which are
more directly related to the neutrino--neutralino
properties~\cite{Hirsch:2000ef}.

The bilinear R--parity violating terms give rise to mixing between
Standard Model and SUSY particles. For example, in these models
neutrinos mix with the neutralinos giving rise to the neutrino masses
and mixings after the diagonalization of a $7 \times 7$ mass matrix
for neutrinos and neutralinos. At the tree level only one neutrino
picks up a mass while the others acquire masses only
radiatively~\cite{Hirsch:2000ef,Diaz:2003as,Hirsch:2004he,Dedes:2006ni}.
Typically the tree-level scale is the atmospheric scale and, at this
approximation, solar neutrino oscillations do not take place.  The
solar mass splitting is ``calculable'' and the solar angle acquires a
meaning only when loops are included. It has been checked that current
neutrino oscillation parameters~\cite{Maltoni:2004ei} can be well
reproduced provided $|\epsilon_i| \ll |\mu|$, where $\mu$ denotes the
SUSY bilinear mass parameter~\cite{Hirsch:2000ef}.

\begin{figure}[t]
  \begin{center}
  \includegraphics[width=9cm]{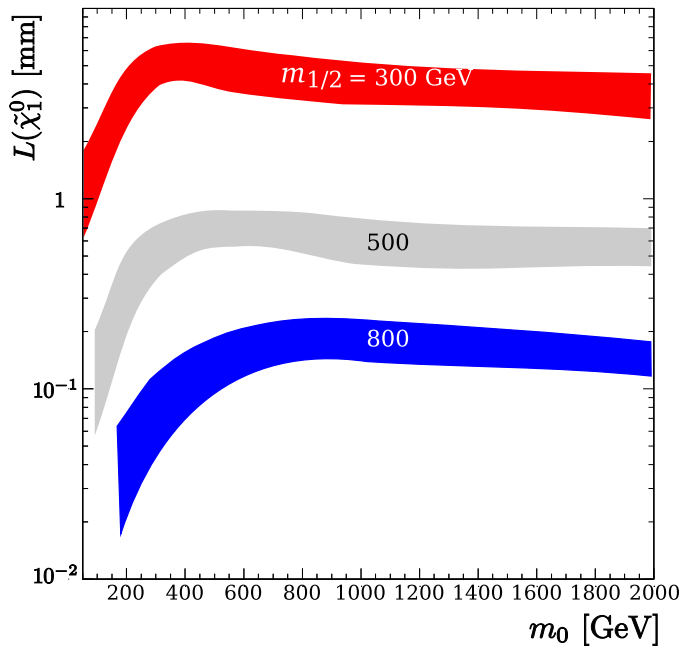}
  \end{center}
  \vspace*{-8mm}
  \caption{ $\tilde\chi_1^0$ decay length versus $m_0$ for $A_0=-100$
    GeV, $\tan\beta=10$, $\mu > 0$, and several values of $m_{1/2}$.
    The widths of the three shaded (colored) bands around
    $m_{1/2}=300,~500,~800$ GeV correspond to the variation of the
    BRpV parameters in such a way that the neutrino masses and mixing
    angles fit the required values~\cite{Maltoni:2004ei} within
    $3\sigma$. }
\label{fig:ldec}
\end{figure}

The LSP lifetime depends upon the SUSY spectrum and the R--parity
violating parameters.  In order to satisfy the neutrino constraints,
the R--parity violating interaction must be feeble, thus leading to a
lifetime large enough to give rise to displaced
vertices~\cite{Porod:2000hv}. For the sake of illustration, we depict
in Figure~\ref{fig:ldec} the $\tilde{\chi}^0_1$ decay length for
$100~\rm{GeV} < m_0 < 2000$ GeV, $300~\rm{GeV} < m_{1/2} < 800$ GeV,
$A_0=-100$ GeV, $\tan(\beta)=10$, and sgn($\mu$)=+1.
We scan randomly all possible BRpV parameter values $\epsilon_i$ and
$\Lambda_i$, and for each SUGRA point in the plot, we accumulate
hundreds of solutions compatible with neutrino oscillation data at 3$\sigma$
level~\cite{Maltoni:2004ei}.
In this way the top (red) band corresponds to $m_{1/2} = 300$ GeV, and
all possible values of $\epsilon_i$ and $\Lambda_i$ compatible with
neutrino mixings and squared mass differences at 3$\sigma$.

It is important to notice that in our BRpV--mSUGRA scheme, for a given
mSUGRA parameter there is a range of variation in the BRpV parameters
$\epsilon_i$ and $\Lambda_i$. While this has no effect in the
production cross sections of supersymmetric states, it affects the LSP
decay length, as illustrated in Fig.~\ref{fig:ldec}. However, the
decay length can vary by not more than $\simeq 30\%$ for a fixed value
of $m_{1/2}$ over the allowed range of the BRpV parameters, except at
small $m_0$ where light scalars play an important part in the LSP
decay.

Apart from generating neutrino masses, neutralino--neutrino mixing
also leads to decay of the LSP into Standard Model particles. In
scenarios where the lightest neutralino is the LSP its main decays are
\begin{itemize}
  
\item {\em leptonic decays:} 

\begin{itemize}

\item $\tilde{\chi}^0_1 
\to \nu \ell^+ \ell^-$ with $\ell =e$, $\mu$ denoted by $\ell \ell$;

\item $\tilde{\chi}^0_1 
\to \nu \tau^+ \tau^-$, called $\tau \tau$;

\item  $\tilde{\chi}^0_1 
\to \tau \nu  \ell$, called $\tau \ell$.

\end{itemize}

\item {\em semi-leptonic decays:}

\begin{itemize}

  \item $\tilde{\chi}^0_1 
\to \nu q \bar{q}$ denoted $jj$;

  \item $\tilde{\chi}^0_1 
\to \tau q^\prime \bar{q}$, called $\tau jj$;

  \item $\tilde{\chi}^0_1 
\to \ell q^\prime \bar{q}$, called $\ell jj$;
    
  \item $\tilde{\chi}^0_1 
\to \nu b \bar{b}$, that we denote by $bb$;

  \item $\tilde{\chi}^0_1 
\to \nu b \bar{b}$, that we denote by $bb$;
\end{itemize}

\item {\em invisible decays:} $\tilde{\chi}^0_1 
\to \nu \nu \nu$.
\end{itemize}


We should mention that many intermediate states contribute to the
above decays, like gauge bosons or scalars. Depending on the spectrum,
some of these final states can be dominated by 2--body decays. For
instance, the decay $\tilde{\chi}^0_1 \to \nu \ell^+ \ell^-$ can be
dominated by the 2--body $\tilde{\chi}^0_1 \to \nu Z$ followed by $Z
\to \ell^+ \ell^-$. Another example is the 2--body mode
$\tilde{\chi}^0_1 \to \nu h$ with $h\to b \bar{b}$. Consequently we
single out the 2--body decays when presenting our results.

\begin{figure}[!h]
  \begin{center}
\includegraphics[width=7cm]{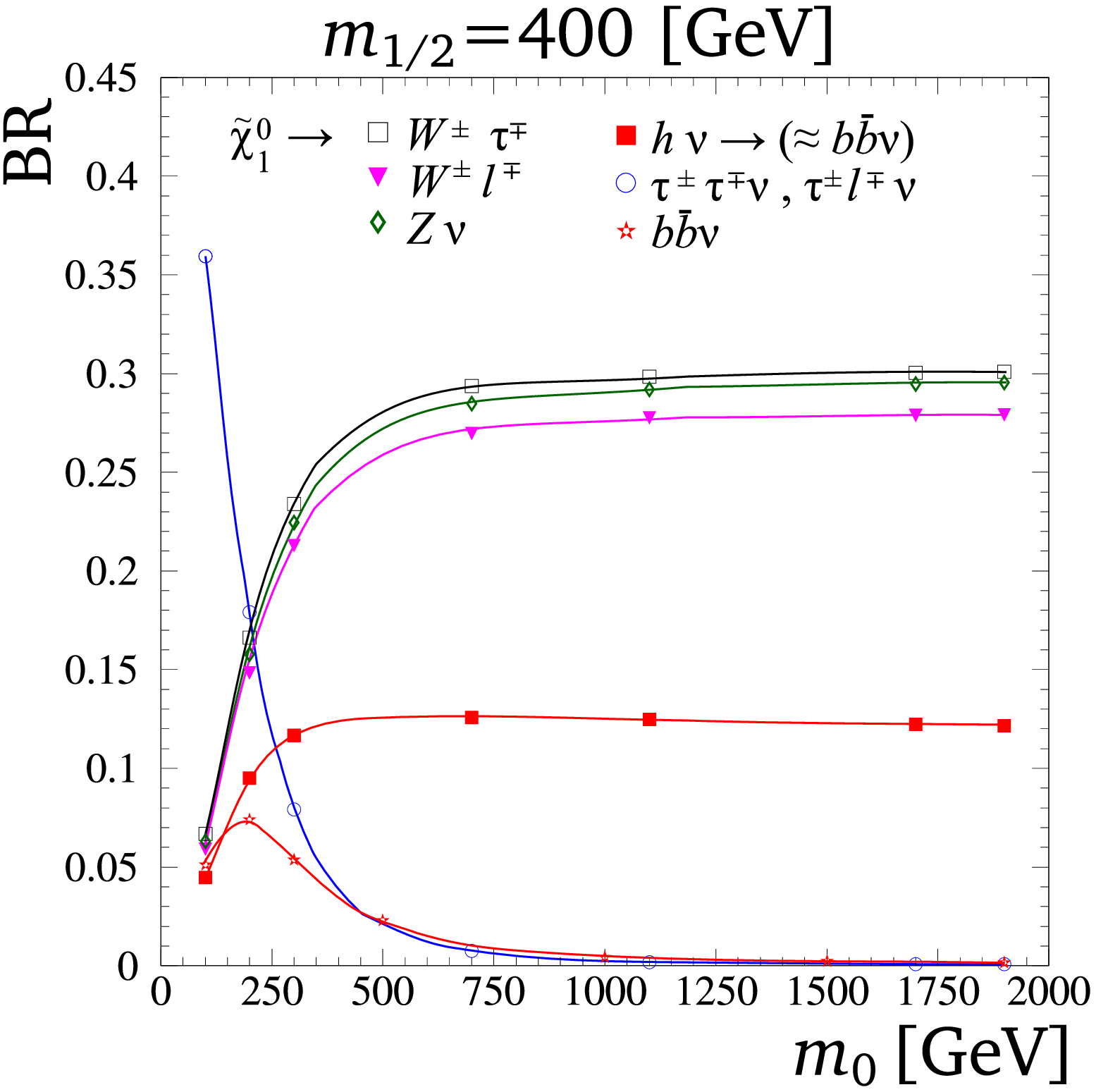}
\includegraphics[width=7cm]{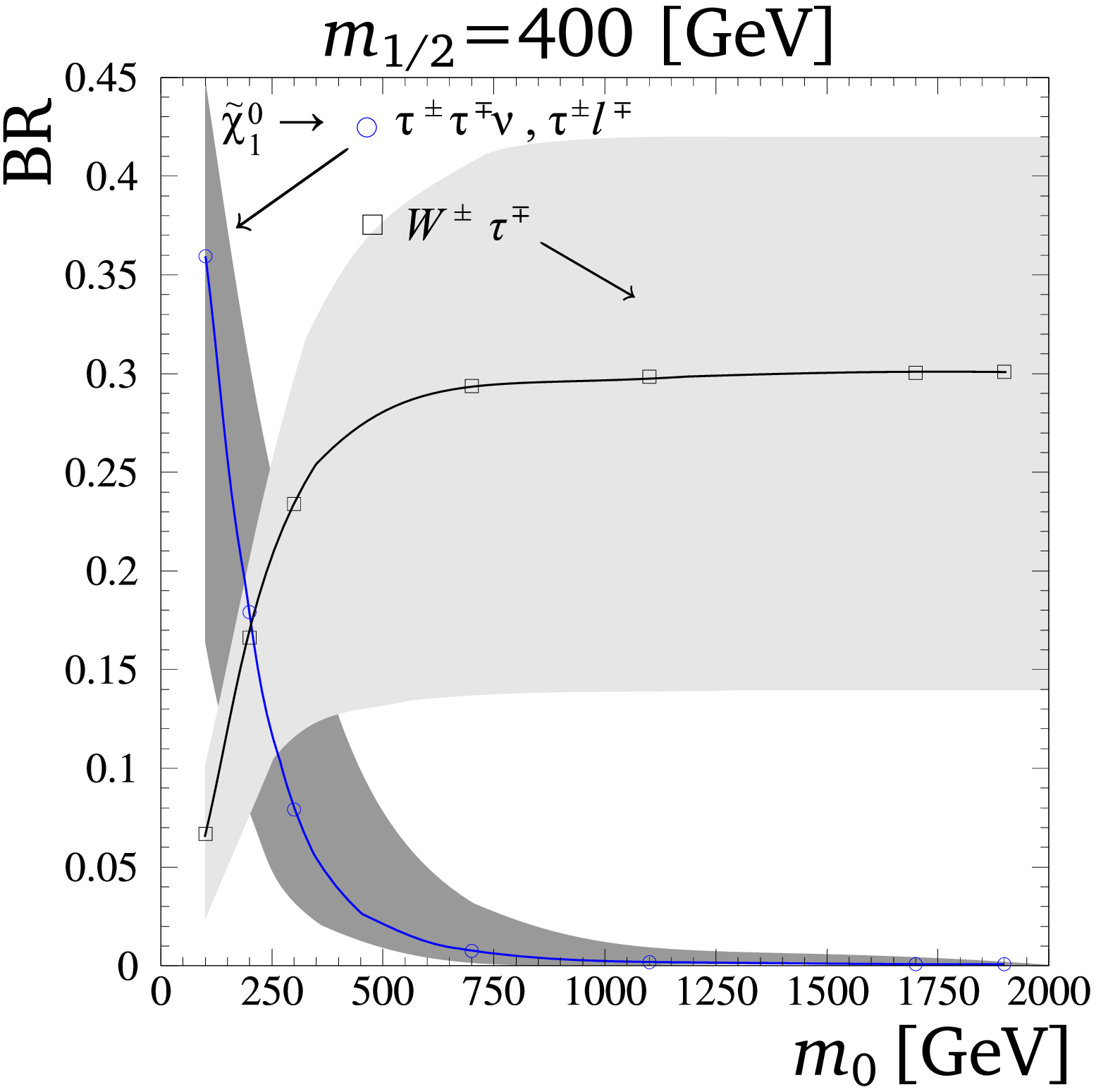}
\vskip .3cm
\includegraphics[width=7cm]{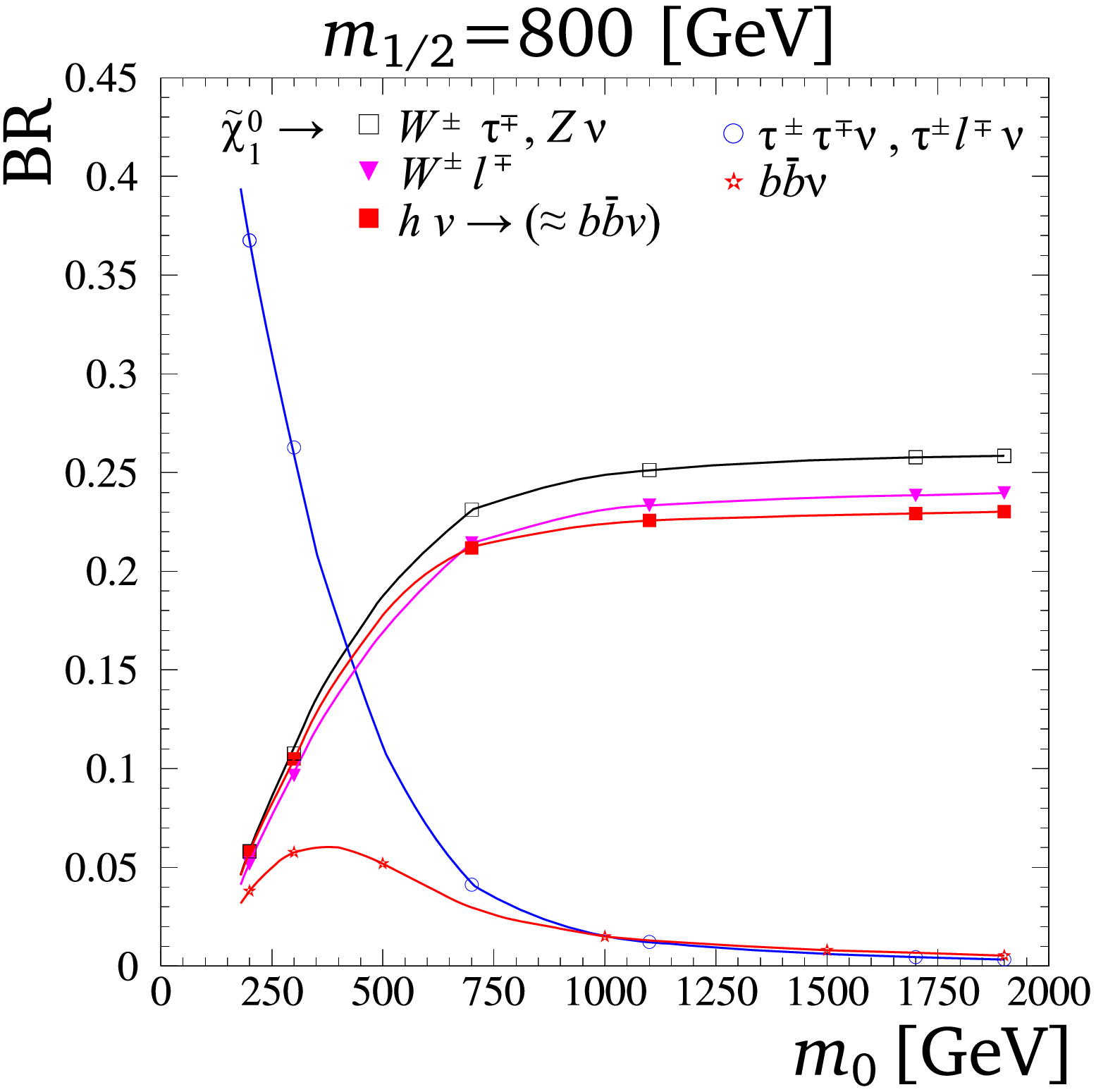}
\includegraphics[width=7cm]{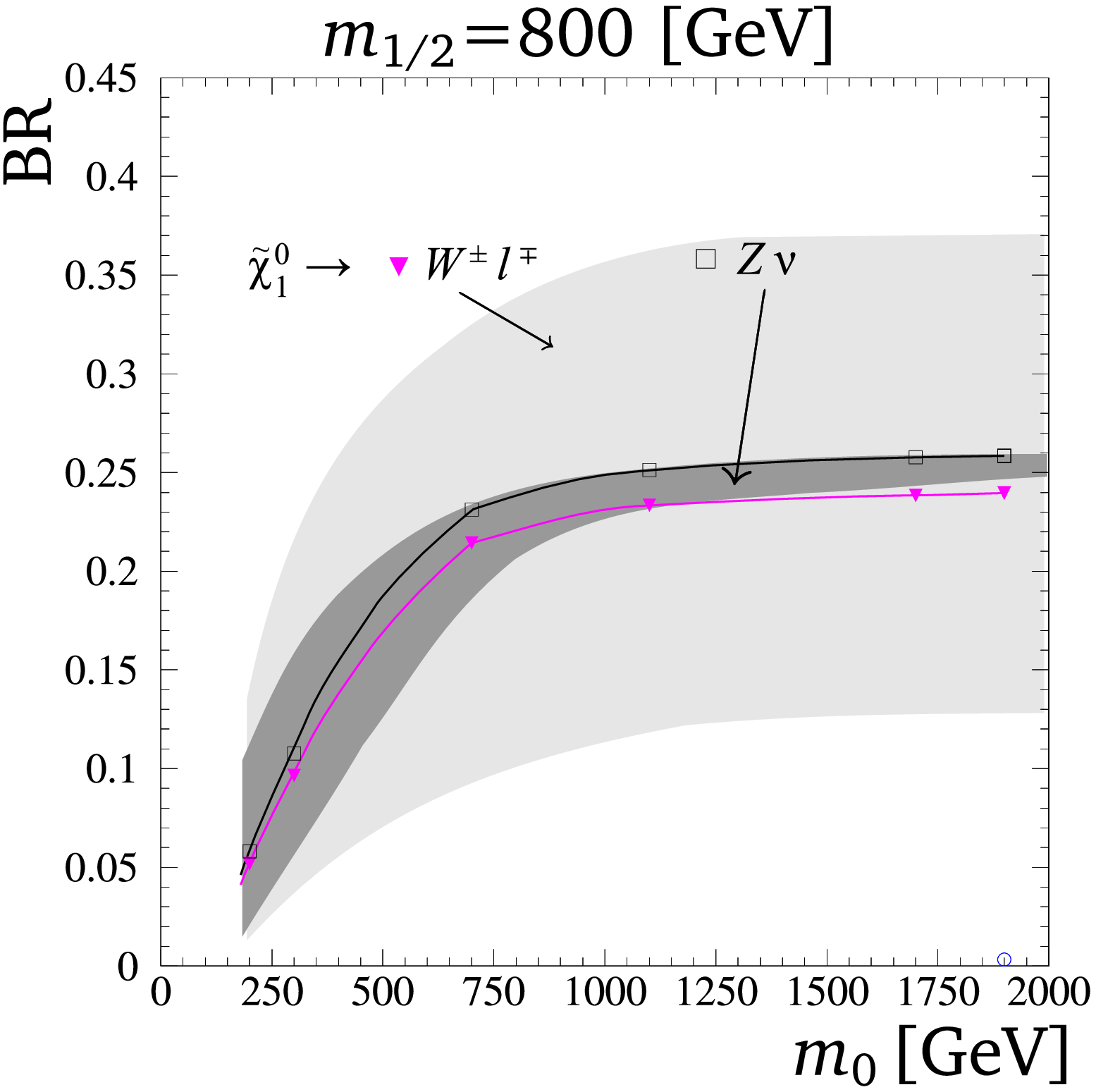}
  \end{center}
  \caption{Lightest neutralino branching ratios as a function of $m_0$
    for $A_0=-100$ GeV, $\tan\beta=10$, and $\mu > 0$. In the upper
    panels we have $m_{1/2} =400$ GeV while the lower ones the
    correspond to $m_{1/2} =800$ GeV. Explanation in text. }
\label{fig:br}
\end{figure}

Figure~\ref{fig:br} presents separately neutralino branching ratios
into two-body and three-body decay final states as a function of $m_0$
for $A_0=-100$ GeV, $\tan\beta=10$ and $\mu > 0$. The top panels
correspond to a fixed value of $m_{1/2}=$ 400~GeV, and the bottom one
correspond to $m_{1/2}=$800~GeV. In the left panels we give branching
ratio predictions for optimized $\epsilon_i$ and $\Lambda_i$ BRpV
parameter values that reproduce the best neutrino oscillation
parameters~\cite{Maltoni:2004ei}.

Recent neutrino oscillation data from reactors (KamLAND) and
accelerators (MINOS) have brought an improved determination of the
solar and atmospheric mass squared splittings~\cite{Maltoni:2004ei}.
On the other hand the solar angle is relatively well determined by
current solar neutrino data.
In contrast, current uncertainties in the determination of the
atmospheric angle remain large and unlikely to improve in the near
future. This translates into a large uncertainty in the expectations
for the LSP decays into gauge bosons, as illustrated in the right
panels in Fig.~\ref{fig:br}.

The light and dark-shaded grey bands in the right panels in
Fig.~\ref{fig:br} indicate, as examples, the uncertainties for
$\tilde{\chi}^0_1 \to W^+ \tau^-$ and $\tilde{\chi}^0_1 \to \nu \tau^+
\tau^-$ respectively (note that similar uncertainties exist for other
channels. For example, $\tilde{\chi}^0_1 \to W^+ \mu^-$ has a similar
spread as $\tilde{\chi}^0_1 \to W^+ \tau^-$ but, to avoid an
over-crowded plot, we have not displayed).
The width of these bands illustrates the effect of varying the BRpV
parameters randomly while satisfying the current neutrino oscillation
data, especially the atmospheric angle, at $3\sigma$.

As one can see, for best-chosen values (left panels) the main LSP
decay mode at small $m_0$ is into lepton pairs including a tau, due to
light scalar tau exchange. In contrast, for large $m_0$ values the
main decays are into charged lepton ($e$, $\mu$ or $\tau$) accompanied
by jets coming from a $W$.

\section{Analysis}
\label{sec:analysis}

In what follows we will describe our two main analysis methods
employing first the conventional SUSY search methods and then taking
advantage of the displaced vertex technique. As already mentioned all
the SUSY particle decays, both R--parity violating as well as
R--parity conserving, are calculated using a generalized version of
the SPheno code~\cite{Porod:2003um}\footnote{A private version can be
  obtained sending an email to W.P.}.  Given a point in SUGRA
parameter space, SPheno searches for a set of $\epsilon_i$ and
$\Lambda_i$ that leads to neutrino masses and mixings compatible with
current neutrino oscillation data.  In general, there is a range of
R--parity violating parameters that satisfy the above neutrino
constraints.  As already mentioned, both SUSY spectra and production
cross sections are well approximated by the conventional R--parity
conserving mSUGRA ones, in view of the smallness of R--parity
violation parameters required to fit the neutrino oscillation data.
However, for consistency, our version of SPheno calculates spectra and
branching ratios including BRpV effects.


We employed PYTHIA version
6.409~\cite{Sjostrand:2000wi,Sjostrand:1993yb} to generate the signal
and backgrounds, inputting the SPheno output in the SLHA
format~\cite{Skands:2003cj}.  Note that we have considered all SUSY
production processes in PYTHIA since they all contain two LSP's in the
decay chain. Moreover, we have assumed that the detector possesses
cells $0.1 \times 0.1$ in $\Delta \phi \times \Delta \eta$ and
included a Gaussian smearing of the energy with $\Delta(E) = 0.5
\times \sqrt{E}$ ($E$ in GeV). Jets were defined using the subroutine
PYCELL with a cone size of $\Delta R = 0.7$. We chose the CTEQ5L
parton distribution functions.

\subsection{Standard searches}
\label{sec:stand-searches}

The LSP decay inside the detector certainly has an important impact in
the supersymmetric particle searches at the LHC. For example, visible
LSP decays reduce the missing transverse energy leading to a weakening
of the standard supersymmetry signals. In order to access the impact
of the LSP decay in the usual SUSY searches we analyzed the LHC
potential for searching for SUSY in the same final states analyzed in
\cite{Baer:2000bs}, {\em i.e.}
\begin{enumerate}

\item {\em Inclusive jets and missing transverse momentum} (denoted by
  {\bf(IN)}): in this case just jets and missing $\sla{p}_T$ are taken into
  account in the analysis;

\item {\em Zero lepton, jets and missing transverse momentum} (denoted by
  $\mathbf{(0\ell)}$): here we consider only events presenting jets and
  missing $\sla{p}_T$ without isolated charged leptons ($e^\pm \,,\, \mu^\pm$);

\item {\em One lepton, jets and missing transverse momentum} (denoted by
  $\mathbf{(1\ell)}$): we include in this final state topology only events
  presenting jets and missing $\sla{p}_T$ accompanied by just one isolated
  charged lepton ($e^\pm \,,\, \mu^\pm$). Notice that final state $\tau$'s
  contribute to this topology only through their leptonic decays;

\item {\em Opposite sign lepton pair, jets and missing transverse momentum}
  (denoted by {\bf (OS)}): this final state is characterized by the presence
  of jets, missing $\sla{p}_T$, and two isolated leptons of opposite charges;

\item {\em Same sign lepton pair, jets and missing transverse momentum}
  (denoted by {\bf (SS)}): here we consider only events presenting jets and
  missing $\sla{p}_T$ accompanied by two isolated leptons of the same charge;

\item {\em Trileptons, jets and missing transverse momentum} (denoted by
  $\mathbf{(3\ell)}$): we classify in this class events that exhibit jets,
  missing $\sla{p}_T$, and three isolated charged leptons;

\item {\em Multileptons, jets and missing transverse momentum} (denoted by
  $\mathbf{(M\ell)}$): this final state presents jets and missing $\sla{p}_T$
  accompanied by more than three isolated charged leptons.

\end{enumerate}

\subsection{Standard event selection}
\label{sec:stand-event-select}

Our analysis of the topologies defined above follows closely the
procedure of~\cite{Baer:2000bs}.  Initially, we applied the acceptance
cuts:

\medskip

\noindent{\bf AC1} We required at least {\bf two} jets in the event with
\begin{equation}
  p^j_T  >  50 \hbox{ GeV} \;\;\; \hbox{and} \;\;\; |\eta_j| < 3  \;.
\end{equation}

\noindent{\bf AC2} In order to suppress a large fraction of the enormous two jet
background, we demanded that the transverse sphericity of the event exceeds
0.2
\begin{equation}
  S_T > 0.2 \; .
\end{equation}


\noindent{\bf AC3} To reduce the background stemming from mismeasured
jets, the missing transverse momentum should not be aligned with any
jet. Therefore, we imposed that the azimuthal angle ($\Delta \varphi$)
between the jets and the missing momentum must comply with
\begin{equation}
  \frac{\pi}{6} \le \Delta \varphi \le \frac{\pi}{2} \; .
\end{equation}

\medskip

In the studies of the different final state topologies we used a floating cut
$E_T^c$ designed to give some optimization of the cuts over the parameter
space~\cite{Baer:2000bs}. We considered $E_T^c = 200$, 300, 400, and 500 GeV.
Given one value of $E_T^c$, we required for all final states that
\begin{equation}
 p_T^{1,2} > E_T^c  \;\;\; \hbox{and} \;\;\;  
 \sla{p}_T > E_T^c  
\label{cuts:1}
\end{equation}
where $ p_T^{1,2}$ stand for the transverse momenta of the two hardest
jets.  These are the only cuts applied for the {\bf IN} topology in
addition to {\bf AC1}--{\bf AC3}. Certainly, we could improve the
reach in this channel, however, this is beyond the scope of our
analysis that aims to access the impact of the LSP decay on the usual
SUSY signals.

For the $\mathbf{0\ell}$ topology we further veto the presence of isolated
leptons with
\begin{equation}
   p_T^\ell > 10 \hbox{ GeV} \;\;\; \hbox{and} \;\;\;  |\eta_\ell| <
   2.5 \; .
\label{cuts:2}
\end{equation}
A charged lepton is considered isolated if the energy deposited in a cone of
$\Delta R < 0.3$ around its direction is smaller than 5 GeV.

The final state $\mathbf{1\ell}$ is obtained imposing (\ref{cuts:1}) and
requiring the presence of only one isolated lepton satisfying
\begin{equation}
  p_T^\ell > 20\hbox{ GeV} \;\;\; \hbox{and} \;\;\;  |\eta_\ell| < 2.5 \; .
\end{equation}
Aiming the reduction of the potentially large $W$ background, we
further imposed that the transverse mass cut
\begin{equation}
  m_T(\sla{p}_T, p^\ell_T) > 100 \hbox{ GeV} \; .
\end{equation}

The {\bf OS} ({\bf SS}) signal events satisfy (\ref{cuts:1}) and exhibit two
isolated leptons with opposite (same) charge. The hardest isolated lepton must
have
\begin{equation}
p_T^\ell > 20 \hbox{ GeV} \;\;\; \hbox{and} \;\;\;  |\eta_\ell| < 2.5 \; .
\label{cuts:3}
\end{equation}
while the second lepton complies with (\ref{cuts:2}).

Trilepton events ($\mathbf{3\ell}$) must pass the cuts (\ref{cuts:1}) and
present three isolated leptons with the hardest one satisfying (\ref{cuts:3})
and the other two respecting (\ref{cuts:2}).  Analogously, multilepton events
$\mathbf{M\ell}$ are obtained adding one or more isolated lepton to the last
topology passing (\ref{cuts:2}).

\subsection{Standard backgrounds}
\label{sec:backgrounds}

The main standard model backgrounds for the supersymmetry discovery
are

\begin{itemize}
  
 \item QCD processes $pp \to jj X$ which have the highest cross section for
  small values of $E_T^c$;
    
 \item $t \bar{t}$ production that leads to final states $WWbb$ which can lead
  to many of the final state topologies analyzed here.

 \item weak gauge boson production in association with jets, denoted
   by $W j$ and $Z j$;

 \item production of electroweak gauge boson pairs $VV$ with $V=Z$ or $W$;

 \item Single top production. Here, we did not include the gluon-$W$ fusion
  contribution since it is not available in PYTHIA.

\end{itemize}

\begin{figure}[th]
  \begin{center}
  \includegraphics[width=7.5cm]{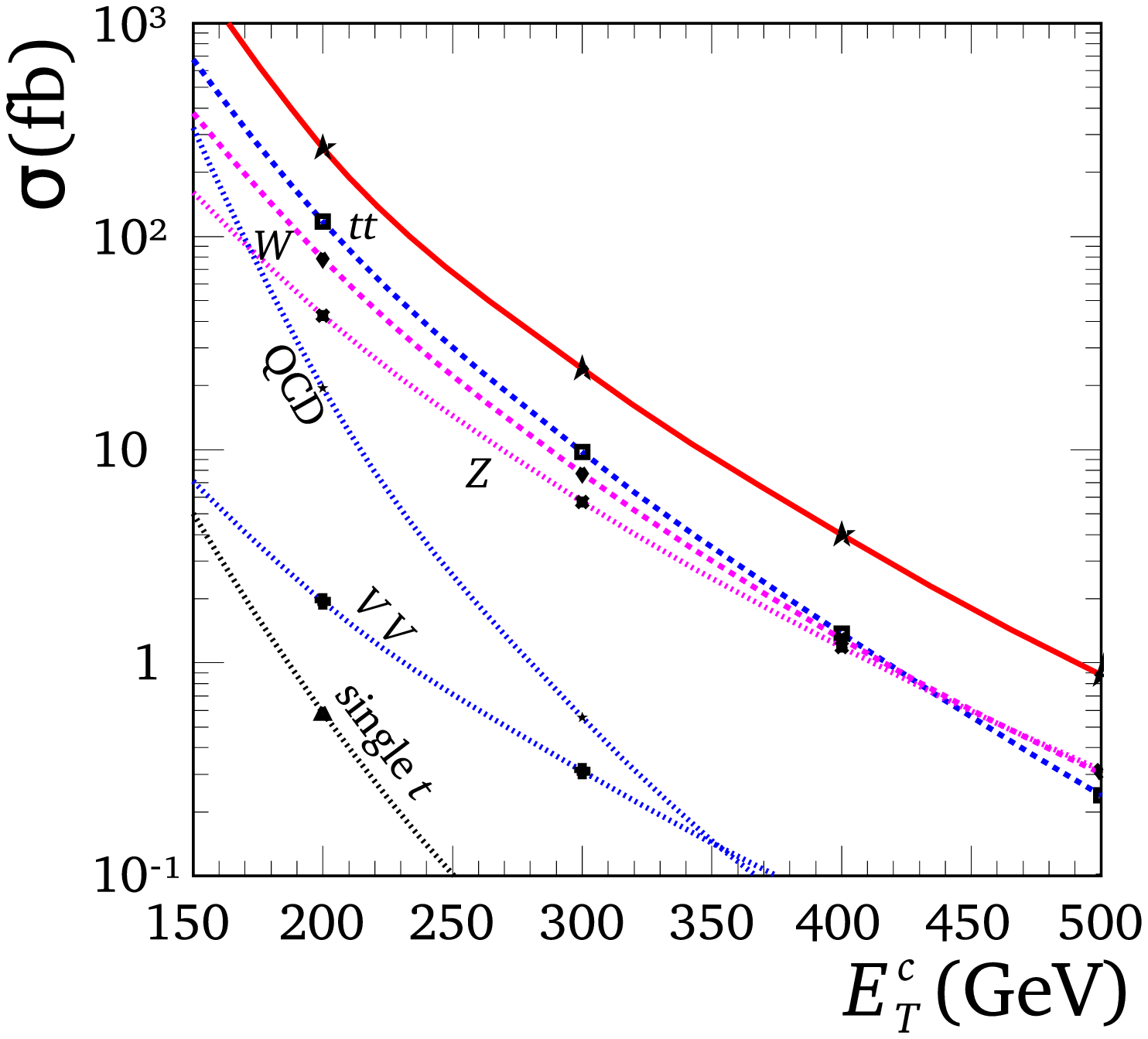}
  \includegraphics[width=7.55cm]{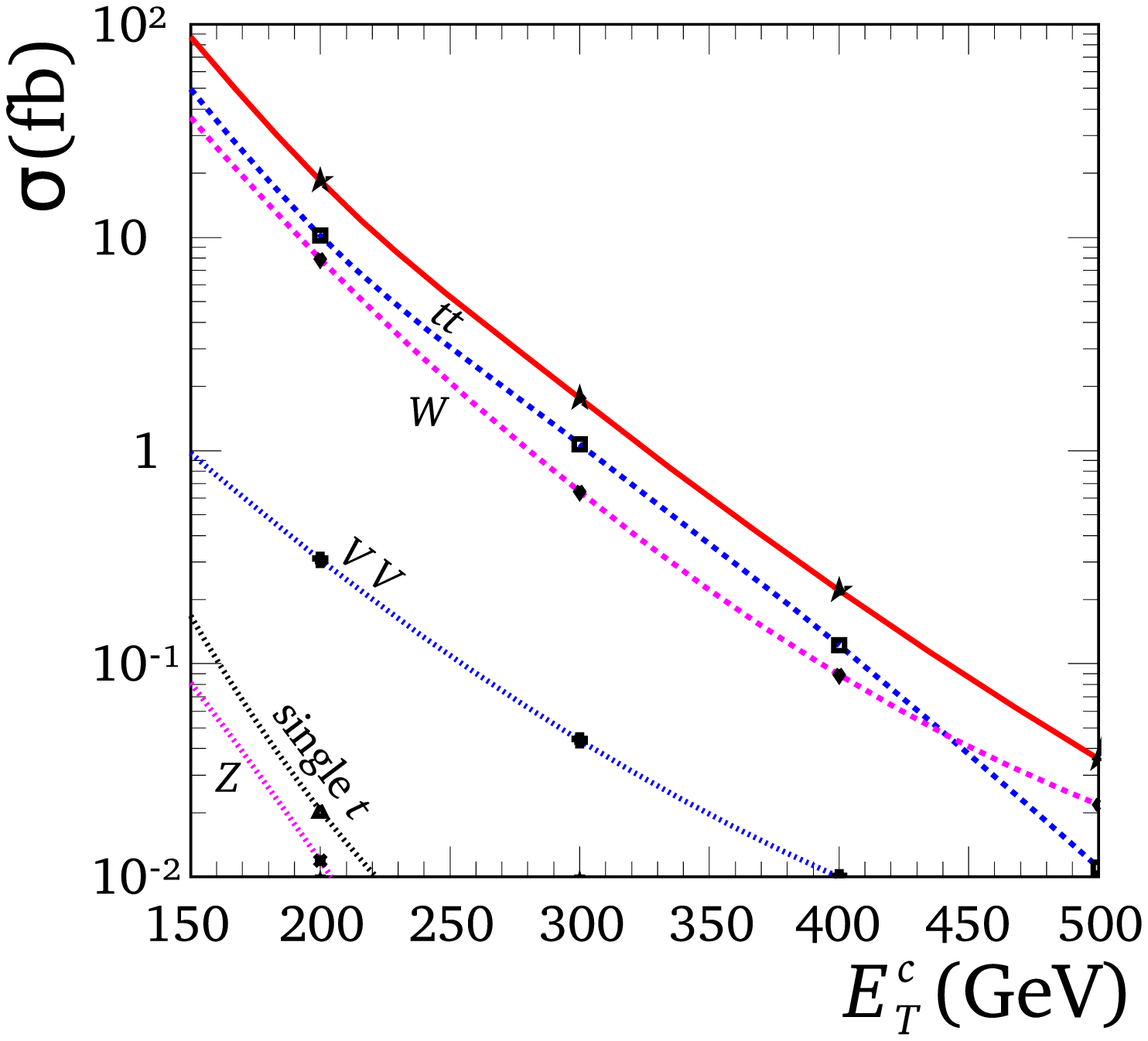}  
  \end{center}
\begin{center}
  \includegraphics[width=7.5cm]{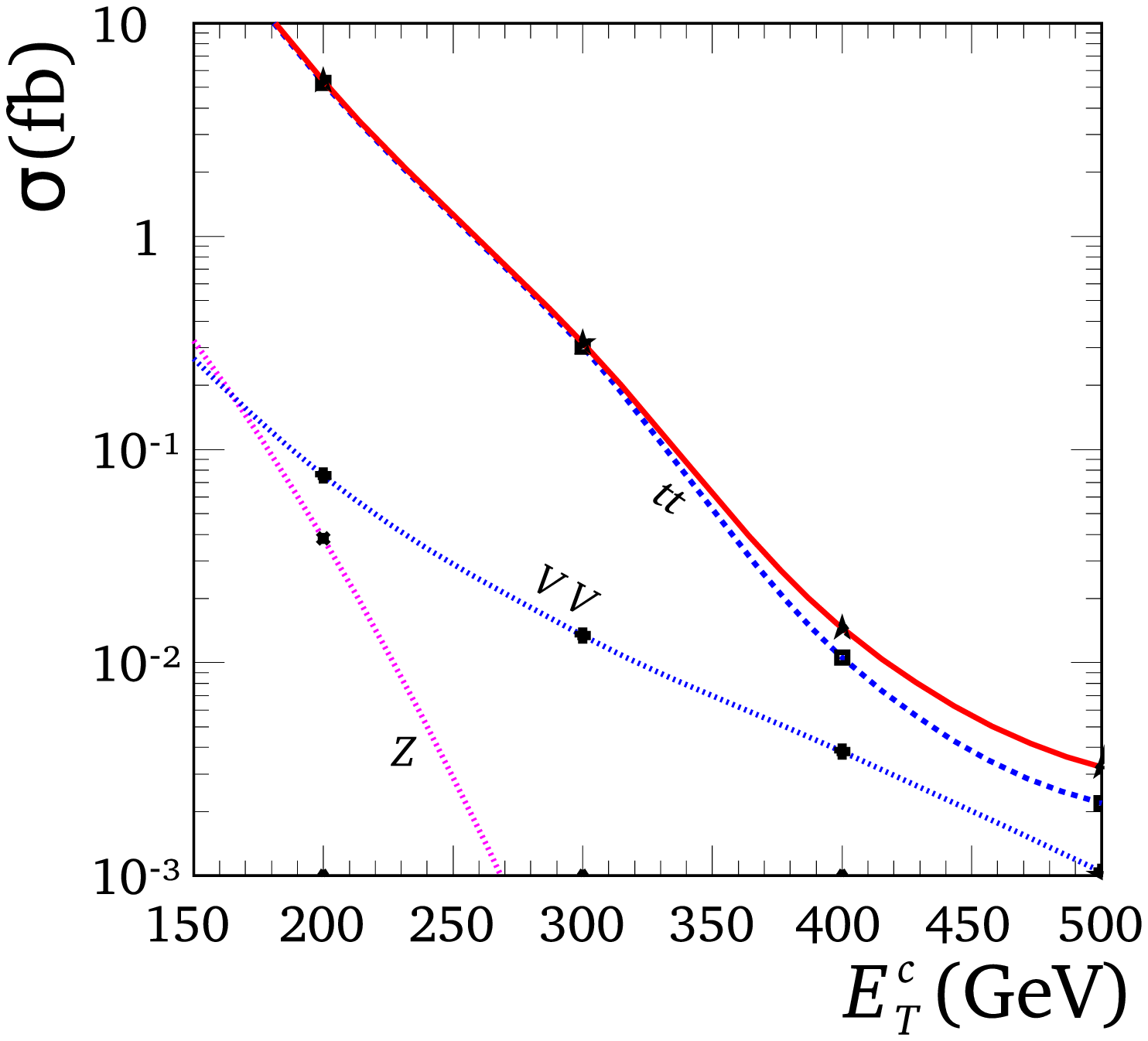} 
\end{center}
\vspace*{-8mm}
  \caption{Composition of the inclusive (top left panel), one lepton
    (top right panel), opposite sign (lower panel) backgrounds.  The
    top (red) solid lines stands for the total backgrounds while the
    blue dashed line corresponds to the $t\bar{t}$ background, the
    magenta dashed lines corresponds to $Wj$, and the magenta dotted
    lines stands for the $Zj$ process. The QCD background is
    represented by the dotted blue line while the dotted black lines
    stands for the single top production and $VV$ is represented by
    the lower dotted blue line at small $E_T^c$.}
\label{fig:back}
\end{figure}

We depict in Figure~\ref{fig:back} the size and composition of the
main SM backgrounds as a function of $E^c_T$ for SUSY searches.
Certainly the SM background for the fully inclusive signal ({\bf IN})
is the largest one. For $E^c_T \simeq 100$ GeV it receives a large
contribution from QCD processes leading to light quarks and gluon,
however, this background decays rapidly with the increase of $E_T^c$.
The main backgrounds for the {\bf IN} topology are $t\bar{t}$ pair
production followed by $Wj$ and $Zj$ productions.  At this point, it
is important to keep in mind that PYTHIA and ISAJET predictions may
differ by large factors since they have different hypothesis in the
fragmentation process, making difficult the comparison of our results
with the ones in reference~\cite{Baer:2000bs}.  For this reason and in
order to have a consistent calculation of signal and backgrounds we
have redone all background calculations, instead of using the results
of this reference.

The $\mathbf{0\ell}$ final state receives a SM backgrounds similar in
size and composition to the ${\bf IN}$ topology, therefore, we do not
show the results for this case. We can see from Figure~\ref{fig:back}
that the bulk of the SM background to the $\mathbf{1\ell}$ topology
originates from $t\bar{t}$ and $Wj$ production over the entire $E_T^c$
range. The SM background for the {\bf OS} background is sizeable for
$E_T^c \lesssim 400$ GeV and it is also dominated by the ubiquitous $t
\bar{t}$ production as shown in Figure~\ref{fig:back}. 
The {\bf SS} background receives its largest contribution from $t
\bar{t}$ pair production with one of them coming from the
semi-leptonic $b$ decay, however, it is much smaller than the previous
ones and negligible for $E_T^c \gtrsim 200$ GeV.  Furthermore, we
verified that the Standard Model processes leading to three or more
isolated leptons are strongly suppressed, so that the searches are
basically background free for the $E_T^c$ range considered in this
paper.

\subsection{Displaced vertex analysis}
\label{sec:displ-vert-analys}

Many of the lightest neutralino decay modes presented in
Section~\ref{sec:model} can be reconstructed by ATLAS and CMS.  We
concentrated our analysis on the $\ell \ell$, $jj$, $\tau jj$, $bb$,
$\tau \tau$, and $\tau \ell$ modes. We did not try to fully identify
these decays modes because we are only interested in the possible
detection of the displaced vertices and not on the further information
that can be obtained from them; this analysis will be the subject of
another work~\cite{prepa}.  For instance, in the decays containing
$\tau$'s, we kept only the one-- and three--prong decays, while we did
not study strategies to differentiate the $bb$ from $jj$ modes. In
order to assure an almost background free search, we required two
reconstructed displaced vertices for each event that pass the requirements
presented below.


In our analysis we assumed a toy detector based on ATLAS technical
proposal~\cite{armstrong1994atp}. In order to be sure that the LSP
decayed away from the primary vertex, we required its decay vertex to
be outside an ellipsoid
\[
      \left ( \frac{x}{\delta_{xy}} \right )^2
   +  \left ( \frac{y}{\delta_{xy}} \right )^2
   +  \left ( \frac{z}{\delta_{z}} \right )^2   = 1 \; ,
\] 
where the $z$-axis is along the beam direction. We were conservative
on the choice of the size of axis of the ellipsoid demanding them to
be five times the ATLAS expected errors in each direction -- that is,
$\delta_{xy} = 20~\mu$m and $\delta_z = 500~\mu$m. We further required
that the LSP decay vertex must be inside the pixel inner detector,
{\em i.e.}  within a radius of $550$ mm and z--axis length of $800$
mm. Moreover, to guarantee a high efficiency in the reconstruction of
the displaced vertices without a full detector simulation, we
restricted our analysis to reconstructed vertex with pseudo-rapidities
$| \eta| < 2.5$.

In LSP decays containing $\tau$'s or $b$ quarks, there might  be
additional displaced vertices present further away from the interaction point.
Therefore, to reconstruct the LSP decay vertices we required that all visible
tracks coming from lightest neutralino decay cascade must cross a sphere
of $10~\mu$m around the LSP vertex.

Relatively long lived SM particles, like $\tau$'s or $B$'s, can also
give rise to displaced vertices as well. However, the Standard Model
physics background can be eliminated by requiring that the set of
tracks defining a displaced vertex must have an invariant mass larger
than 20~GeV. This way the displaced vertex signal passing all the
above cuts is essentially background free, except for possible
instrument backgrounds which are beyond the scope of this analysis.

At this point we must make sure that the events presenting LSP
displaced vertices pass at least of the triggers to guarantee that
they will be properly recorded.  In order to mimic the triggers used
by the LHC collaborations, we accept events passing at least one of
the following requirements:
\begin{itemize}

\item the event has one isolated electron with $p_T > 20$ GeV and $|\eta|<2.5$;

\item the event has one isolated muon with $p_T > 6$ GeV and $|\eta|<2.5$;

\item the event has two isolated electrons with $p_T > 15$ GeV;

\item the event has one jet with $p_T> 100$ GeV;

\item the event has missing transversal energy in excess of $100$ GeV.

\end{itemize}


\section{ Results}
\label{sec:results}

\subsection{Standard Analysis}
\label{sec:standard-analysis}

We estimated the SUSY discovery potential of the LHC in the {\bf IN},
$\mathbf{0\ell}$, $\mathbf{1\ell}$, $\mathbf{3\ell}$, $\mathbf{M\ell}$, {\bf
  OS}, and {\bf SS} channels. We analyzed each channel independently, not
trying to combine the outcomes of the different final state topologies.
Certainly combining different topologies enhances the LHC reach for SUSY.  For
the sake of definiteness we defined that a point in the parameter space of our
model can be observed in a given channel if there is a choice of $E_T^c$ for
which either the signal leads to $5\sigma$ departure from the background
expectation, where this is not vanishing, or 5 events in regions where the SM
background vanishes.  We present our results in the $m_{1/2} \otimes m_0$
plane for $\tan\beta=10$, $\mu>0$, $A_0 = -100$ GeV, and integrated
luminosities of 10 and 100 fb$^{-1}$.

\begin{figure}[th]
  \begin{center}
 \includegraphics[width=7.75cm]{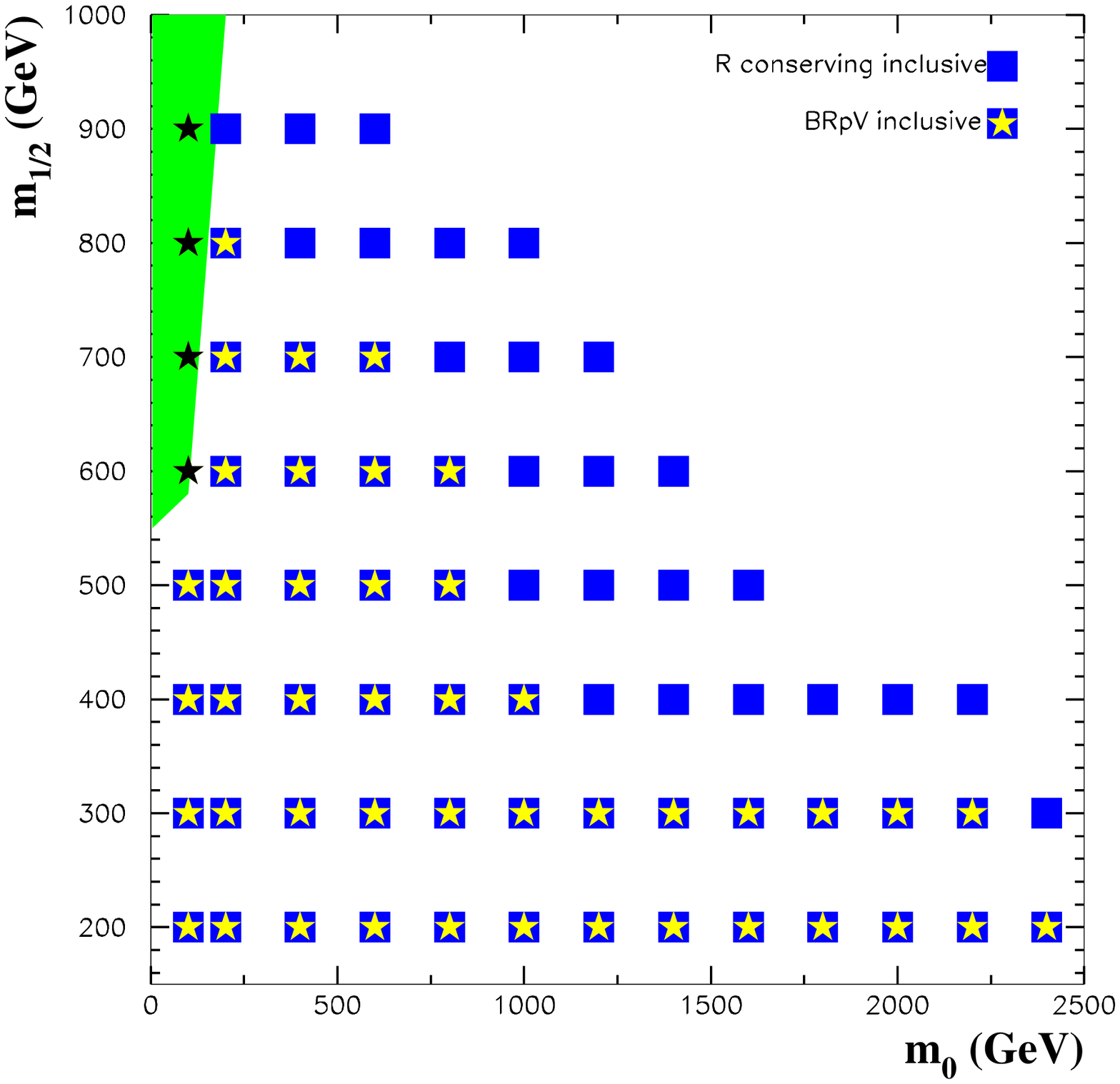}
 \includegraphics[width=7.75cm]{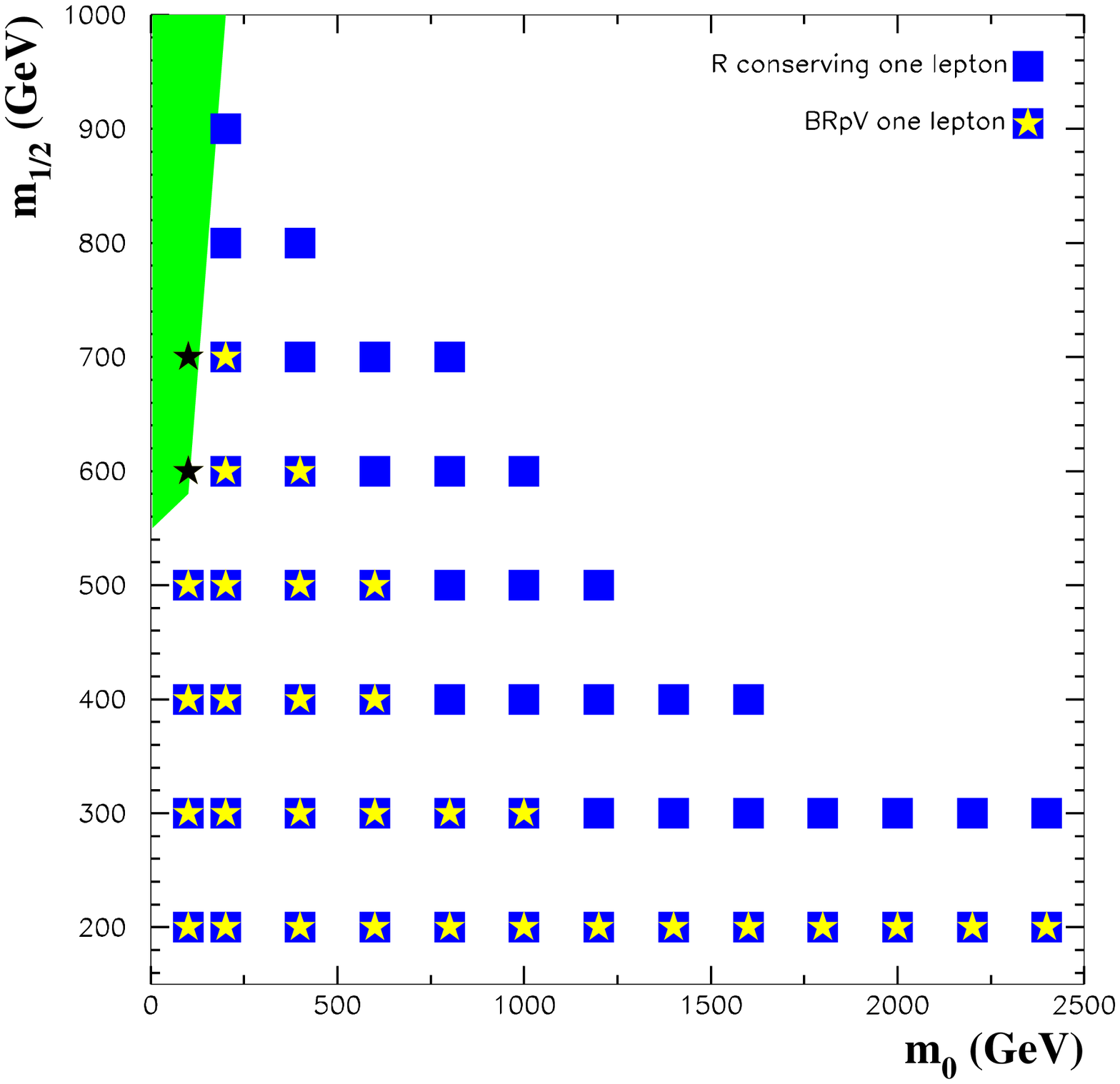}
 \end{center}
 \caption{LHC discovery reach in the inclusive channel (left panel)
   and in the one lepton channel (right panel). We assumed
   $\tan\beta=10$, $\mu>0$, $A_0 = -100$ GeV, and an integrated
   luminosity of 10 fb$^{-1}$.  The points leading to visible signal
   at the LHC are marked by a square for (R-parity-conserving) mSUGRA
   and by a star for the BRpV--mSUGRA case. The (green) shaded area
   indicates the region where the stau is the LSP. 
   Points already excluded by LEP and Tevatron searches lie below the
   $m_{1/2}$ values depicted in this figure.}
\label{fig:inclu10}
\end{figure}

\begin{figure}[th]
  \begin{center}
  \includegraphics[width=7.75cm]{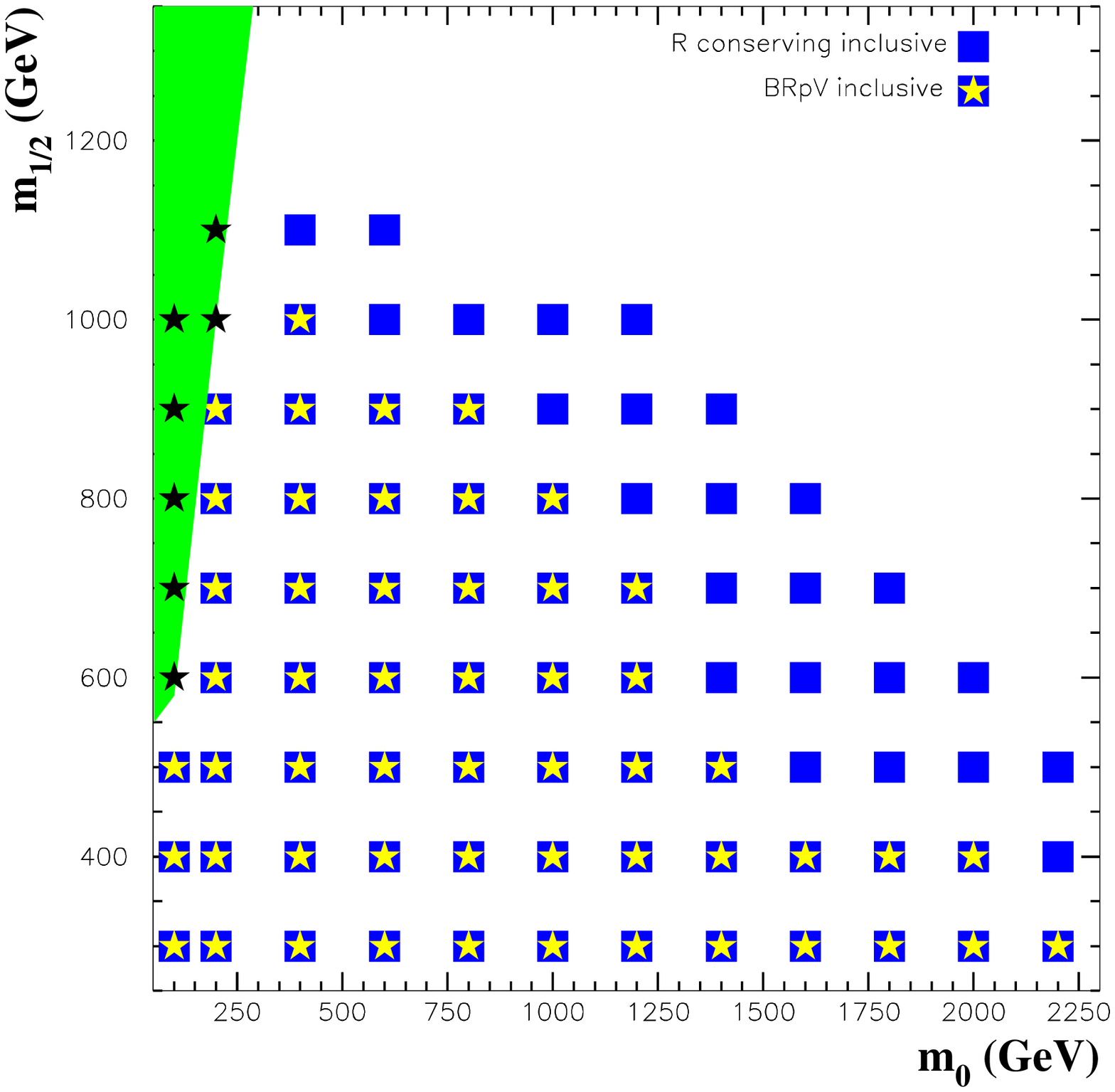}
  \includegraphics[width=7.75cm]{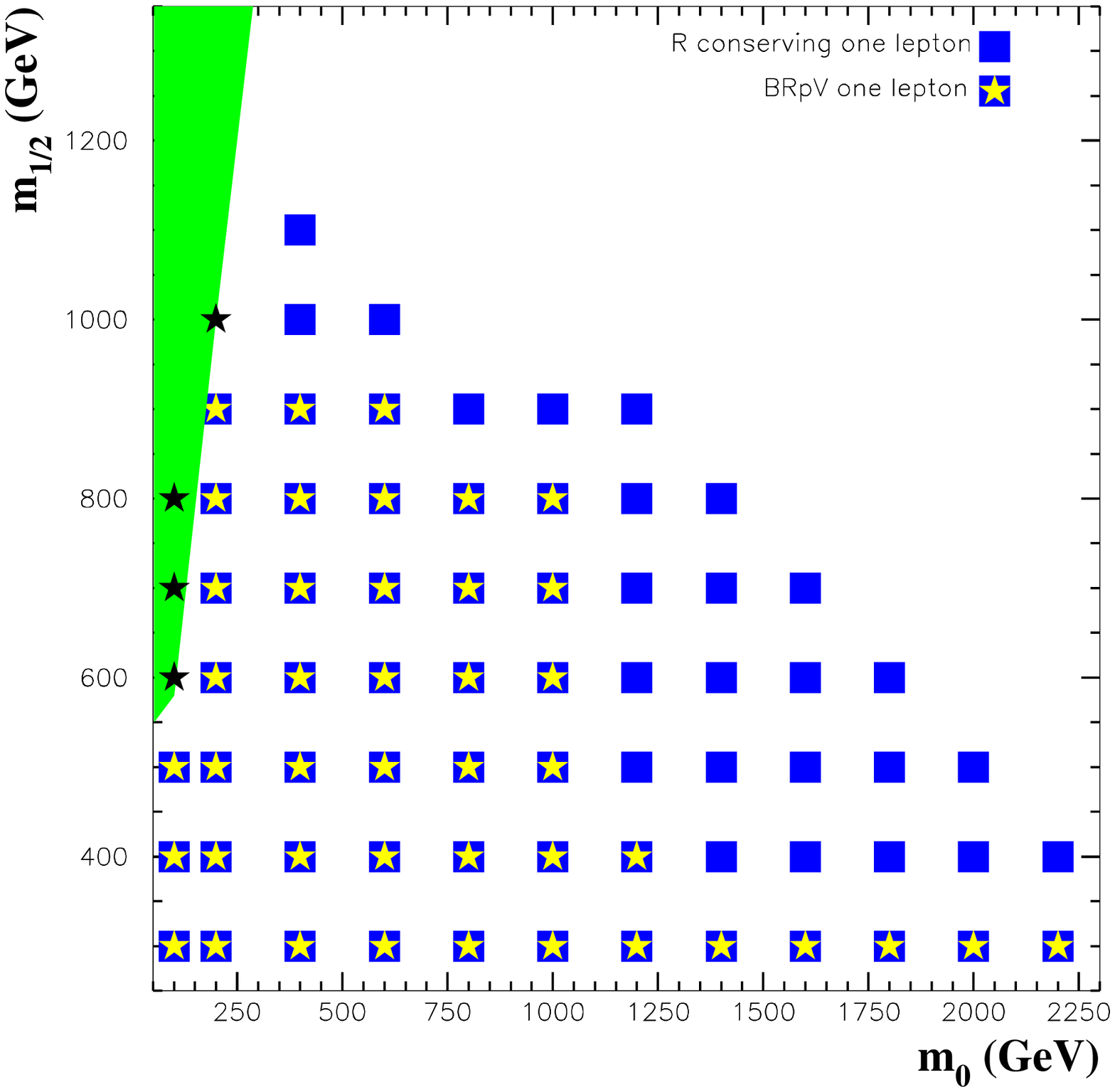}  
  \end{center}
  \vspace*{-8mm}
  \caption{Same as Fig. \ref{fig:inclu10} but for an integrated
    luminosity of $100$ fb$^{-1}$.  }
\label{fig:inclu100}
\end{figure}

We depict in Figure \ref{fig:inclu10} the LHC discovery potential in
the inclusive (left panel) and one lepton (right panel) channels
assuming R--parity conservation as well as bilinear R--parity
violation for an integrated luminosity of 10 fb$^{-1}$; the results
for 100 fb$^{-1}$ are shown in Figure \ref{fig:inclu100}.
Current direct search limits coming from LEP and Tevatron do not
appear in this plot since these constrain lower values of $m_{1/2}$
than the ones displayed, highlighting the increased coverage of the
LHC.
From the left panel of Fig.~\ref{fig:inclu10} one can see how bilinear
R--parity violation reduces the reach in the inclusive channel for a
given value of $m_0$ with the signal depletion growing as $m_0$
increases.  Basically, the decay of the LSP in BRpV--mSUGRA reduces the
available missing transverse energy making it harder for the signal to
pass the cuts, and consequently, stands out from the SM background.
This effect is specially important at moderate and high $m_0$ where
the fraction of decays producing neutrinos diminishes; see
Fig.~\ref{fig:br}.
Moreover, the shaded area indicates the region where the stau is the
LSP. As already discussed, this region is excluded if R--parity is
conserved but not in our BRpV--mSUGRA model.  We note that the inclusive
signal is rather strong in this area.

The reach in the $1\ell$ channel is also presented in
Fig.~\ref{fig:inclu10}, right panel, for both cases, with and without
R--parity conservation. Clearly from Fig.~\ref{fig:inclu10} one can
see that, as expected, the reach in this channel is smaller than the
one in the inclusive channel with or without R--parity conservation
since it diminishes as the slepton masses ($m_0$) increase.  In this
channel, the suppression of the signal is larger than the inclusive
case when the R--parity violating interactions are present, being
visible only at small $m_0$ ($m_{1/2}$) in the BRpV scenario. This
suppression of the one--lepton BRpV signal is due to the reduction of
the missing transverse energy at large $m_0$, while at small $m_0$ it
originates from the production of further leptons in the LSP decay as
well. Remember that in this channel we require one and only one
isolated charged lepton.

The effect of a larger data sample can be seen in Fig.\ \ref{fig:inclu100}
where we assumed an integrated luminosity of 100 fb$^{-1}$. The presence of
BRpV reduces the signal reach in the inclusive and one-lepton channels when
compared to the R--parity conserving case. Moreover, at small and moderate
$m_0$ the inclusive and one-lepton channels lead to similar reach in our BRpV
model, however, the inclusive signal has a larger reach for $m_0 \gtrsim 1$
TeV.

\begin{figure}[th]
  \begin{center}
  \includegraphics[width=7.75cm]{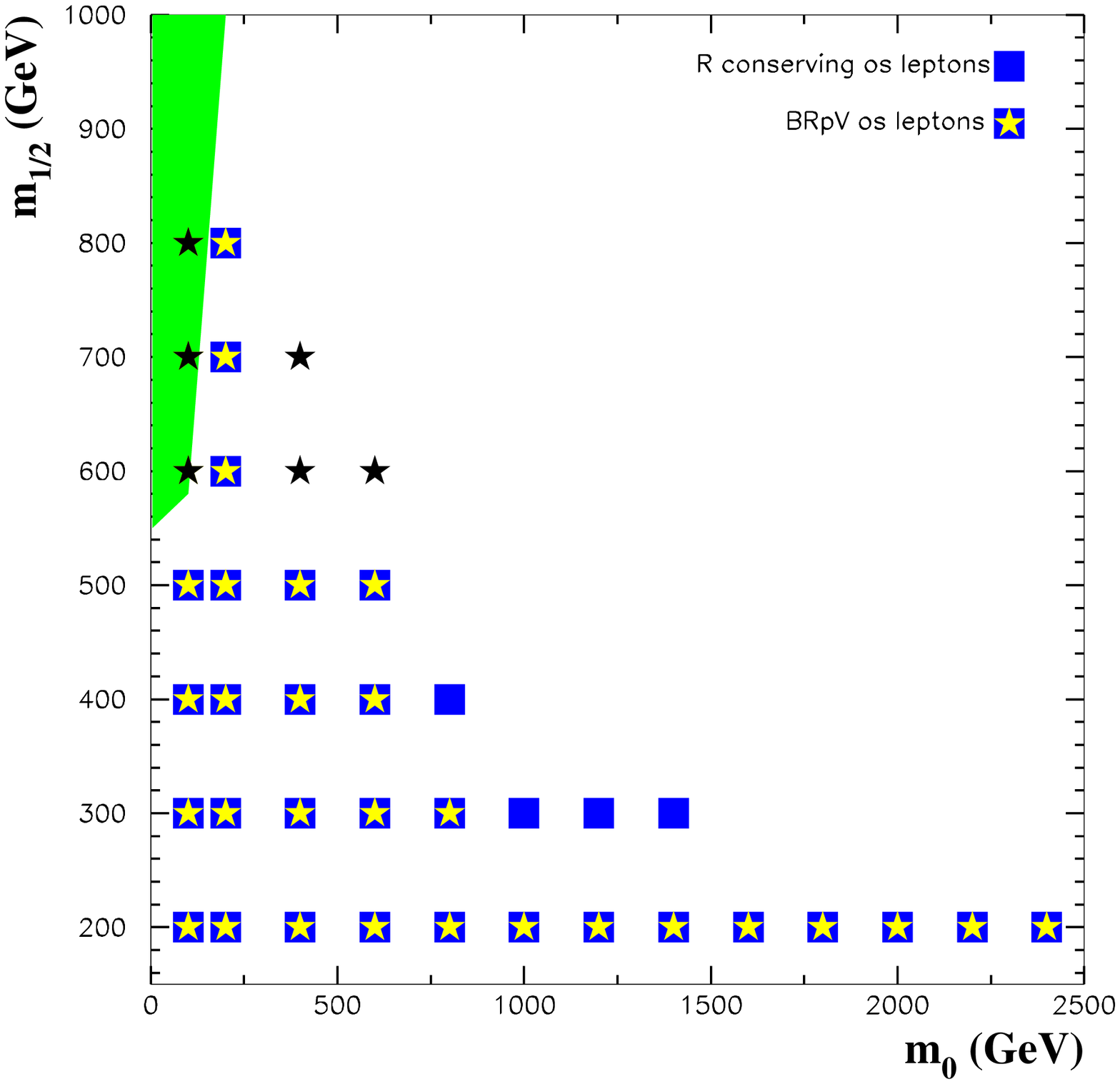}
  \includegraphics[width=7.75cm]{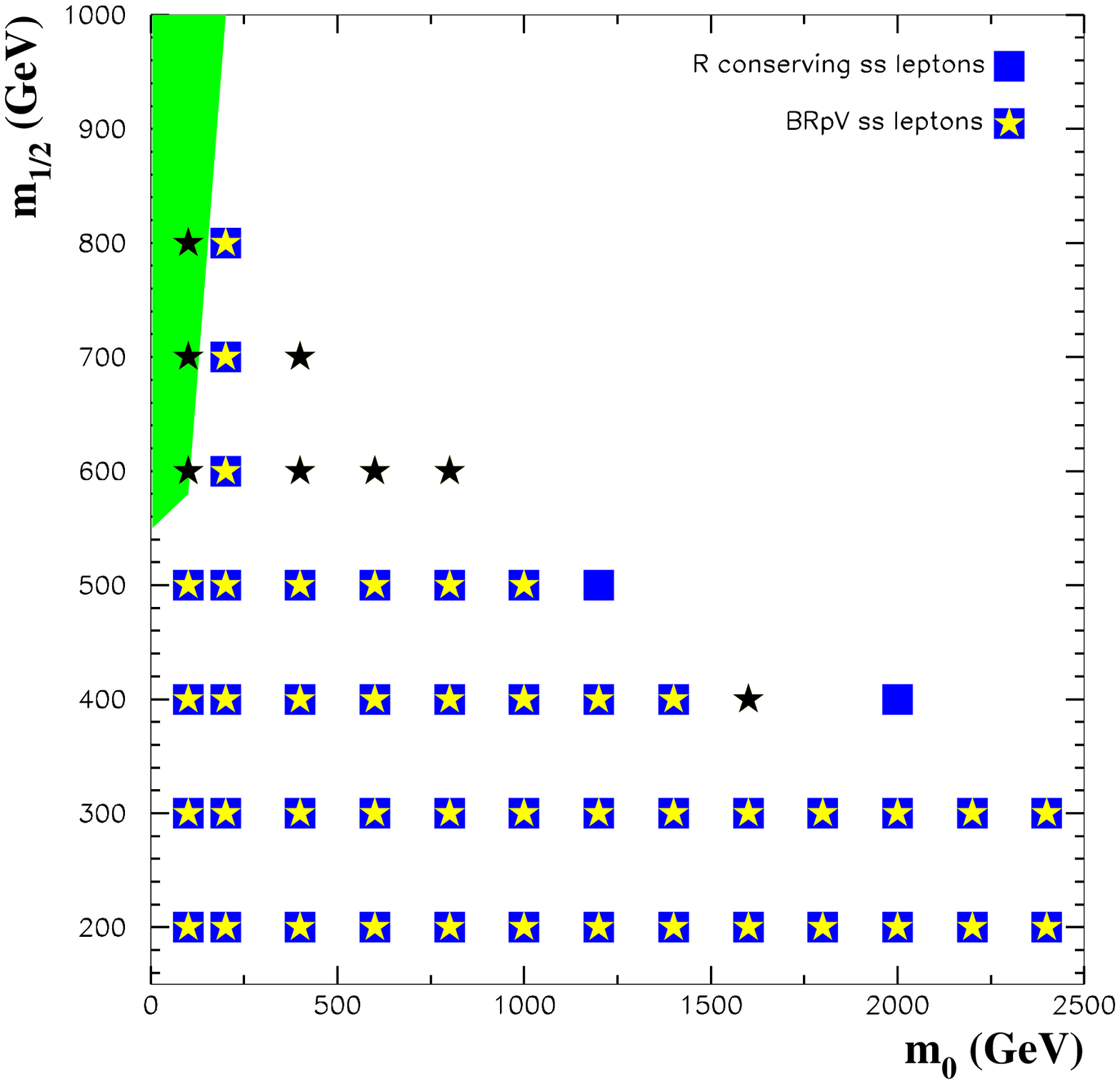}
  \end{center}
  \vspace*{-8mm} \caption{LHC discovery reach using opposite sign
    leptons (left panel) and same sign leptons (right panel) for the
    parameters used in Fig.~\ref{fig:inclu10}.}
\label{fig:di:os10}
\end{figure}

\begin{figure}[th]
  \begin{center}
  \includegraphics[width=7.5cm]{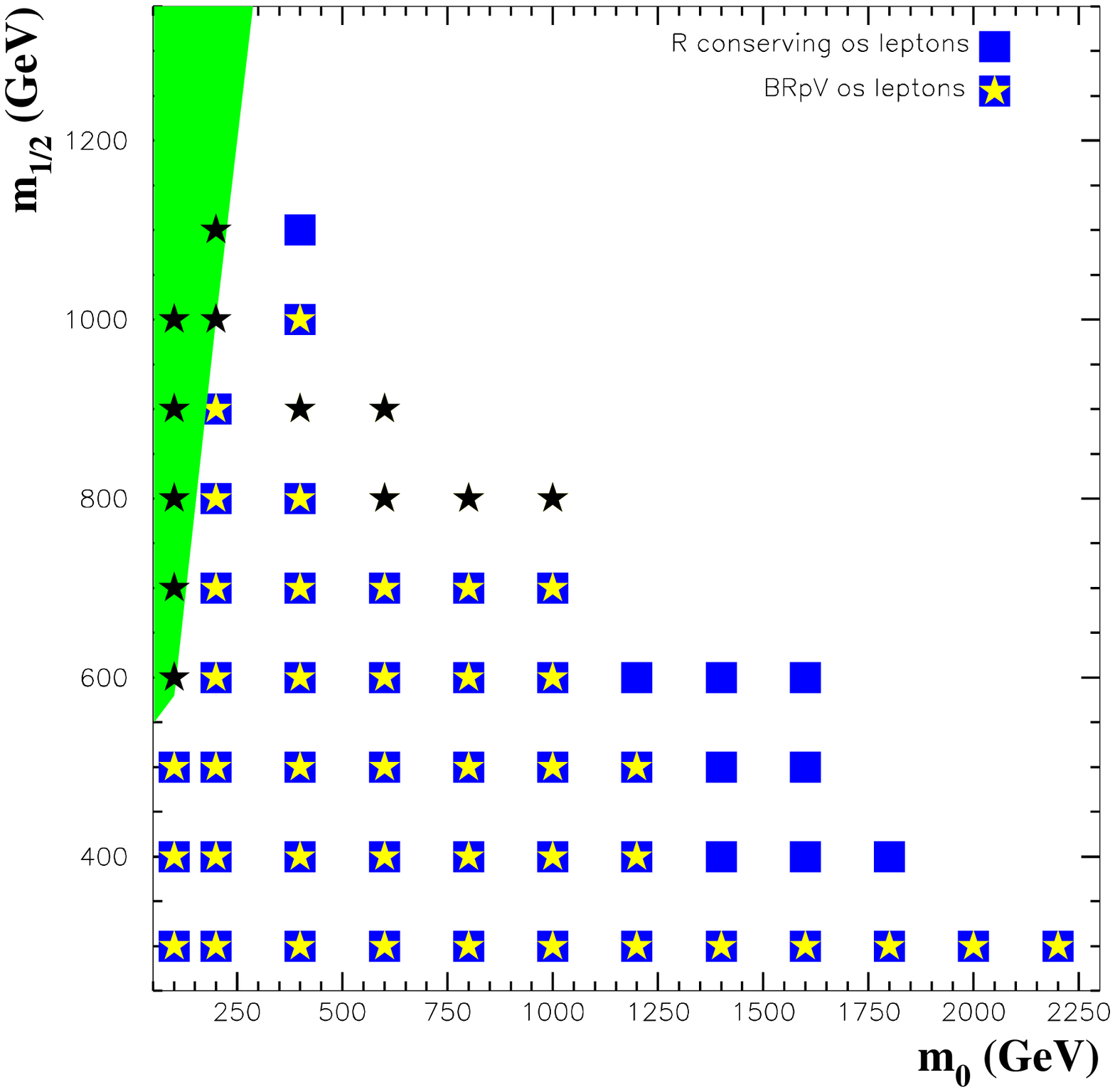}
  \includegraphics[width=7.5cm]{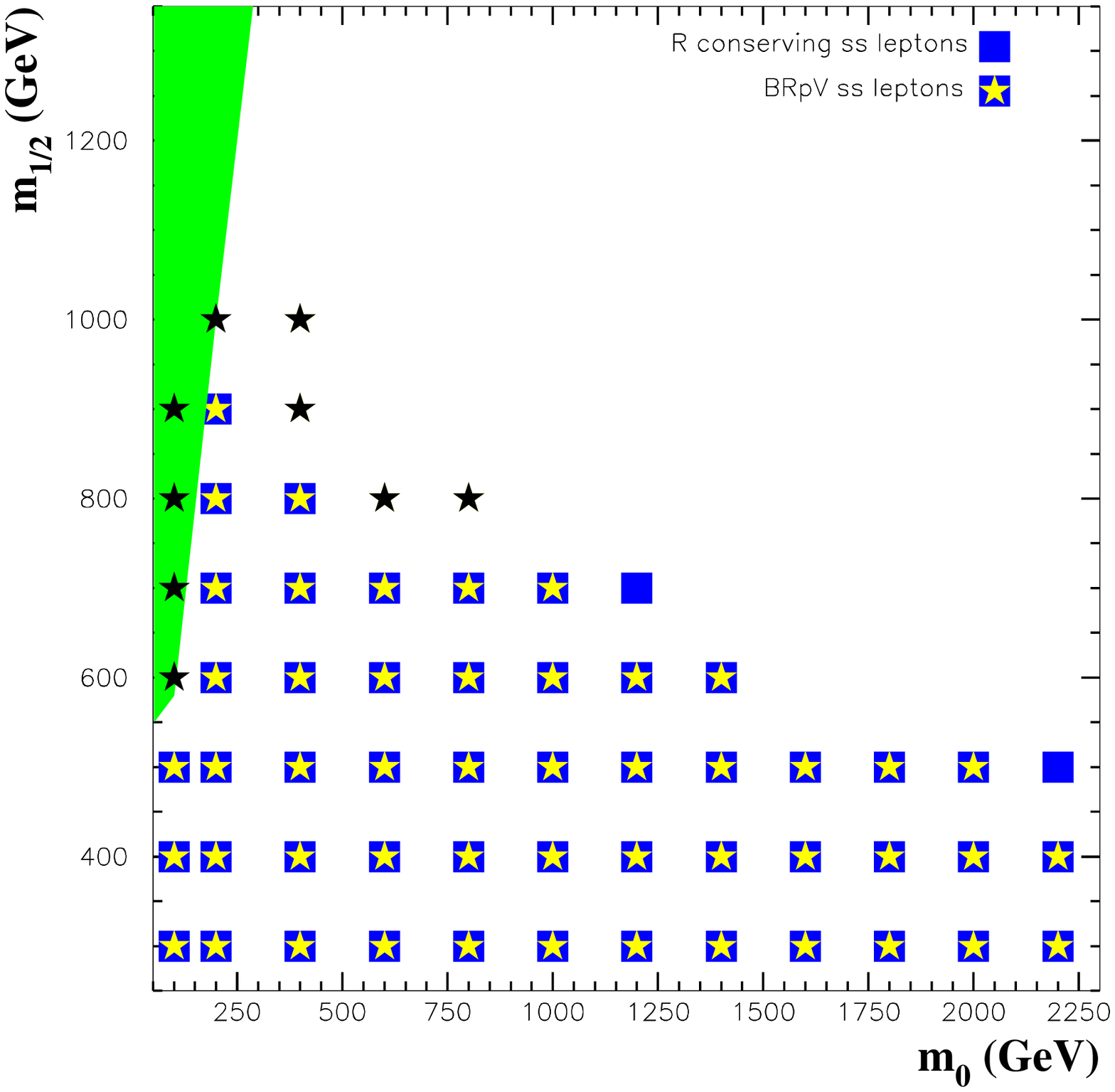}
  \end{center}
  \vspace*{-8mm}
  \caption{Same as Fig. \ref{fig:di:os10} but for an integrated
    luminosity of $100$ fb$^{-1}$.  }
\label{fig:di:os100}
\end{figure}

In Figures~\ref{fig:di:os10} and \ref{fig:di:os100} we present the LHC
discovery potential in the opposite ({\bf OS}) and same-sign ({\bf
  SS}) di-lepton signals ({\bf OS} (left) and {\bf SS} (right))
with/without R--parity conservation for integrated luminosities of 10
and 100 fb$^{-1}$ respectively.
Clearly, the presence of R--parity violation extends the LHC reach in
this channel at small $m_0$ with respect to the R--parity conserving
scenario. On the other hand the coverage at large $m_0$ is similar in
both cases.  Again, this extended reach in the BRpV case is due to the
presence of leptons coming from the LSP decay mediated by light
sleptons.  In the BRpV scenario the {\bf SS} channel provides the
largest reach at moderate and large $m_0$ improving the discovery
potential while its reach is similar to the all inclusive channel for
BRpV at small $m_0$.  The effect of having a larger data sample, for
instance 100 fb$^{-1}$, is shown in Fig.\ \ref{fig:di:os100}, where we
can see that the {\bf SS} signal is more suitable than the fully
inclusive signal to search for BRpV models at large $m_0$.

\begin{figure}[th]
  \begin{center}
\includegraphics[width=7.5cm]{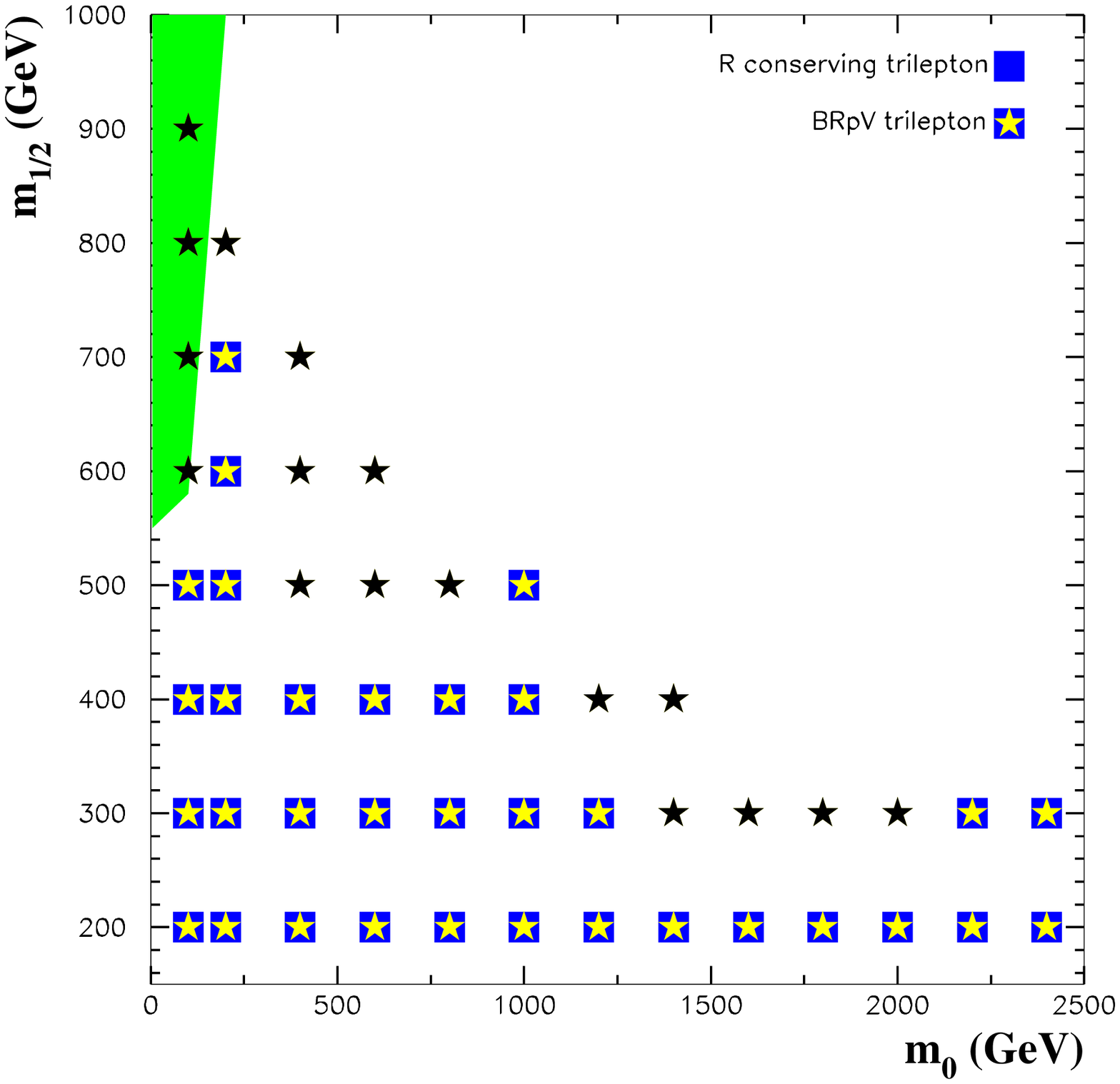}
\includegraphics[width=7.5cm]{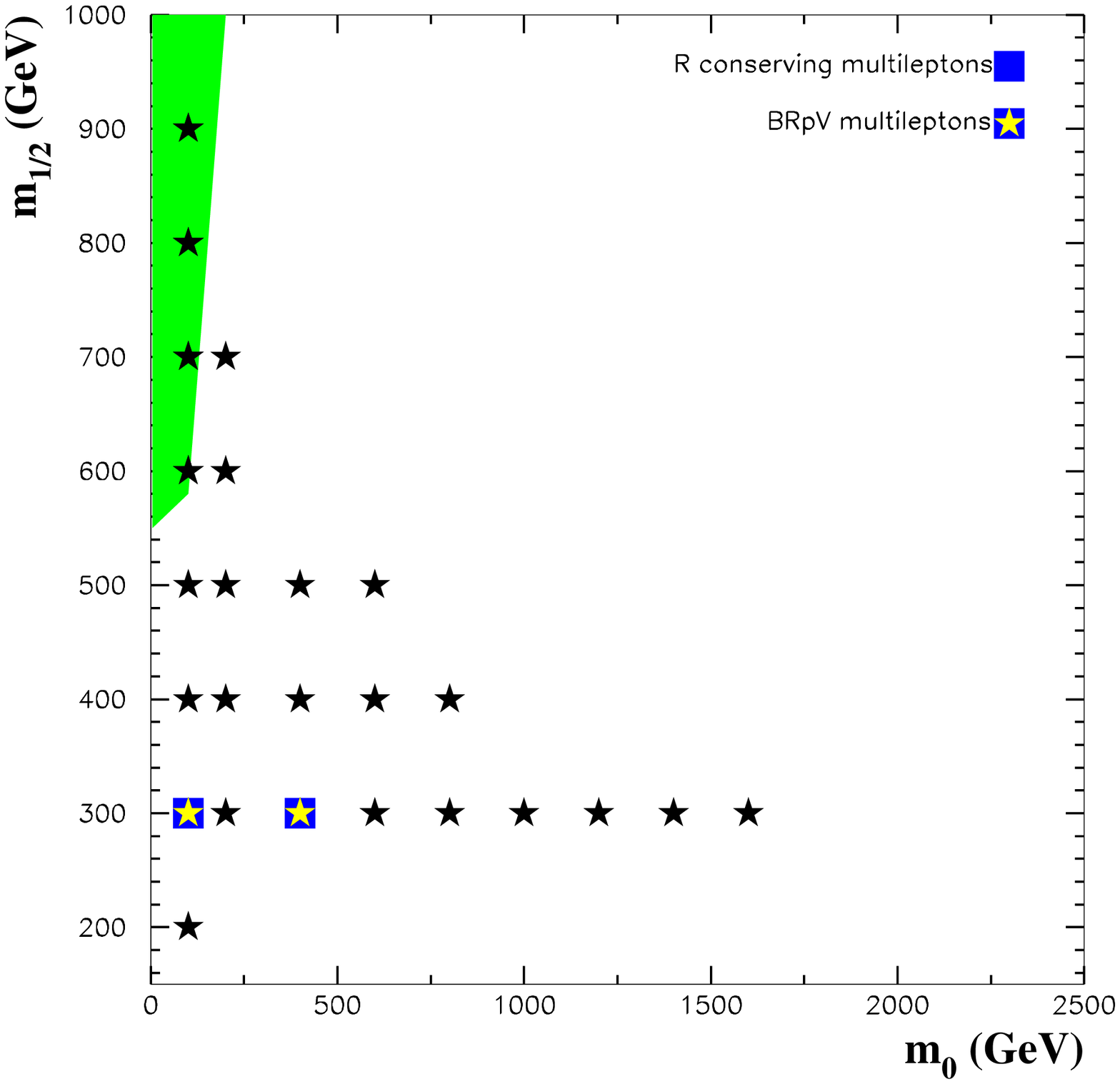}
  \end{center}
  \vspace*{-8mm}
  \caption{LHC discovery potential in the three lepton channel (left
    panel) and the multilepton one (right panel) for the parameters
    used in Fig.~\ref{fig:inclu10}.}
\label{fig:di:3lep10}
\end{figure}

\begin{figure}[th]
  \begin{center}
 \includegraphics[width=7.5cm]{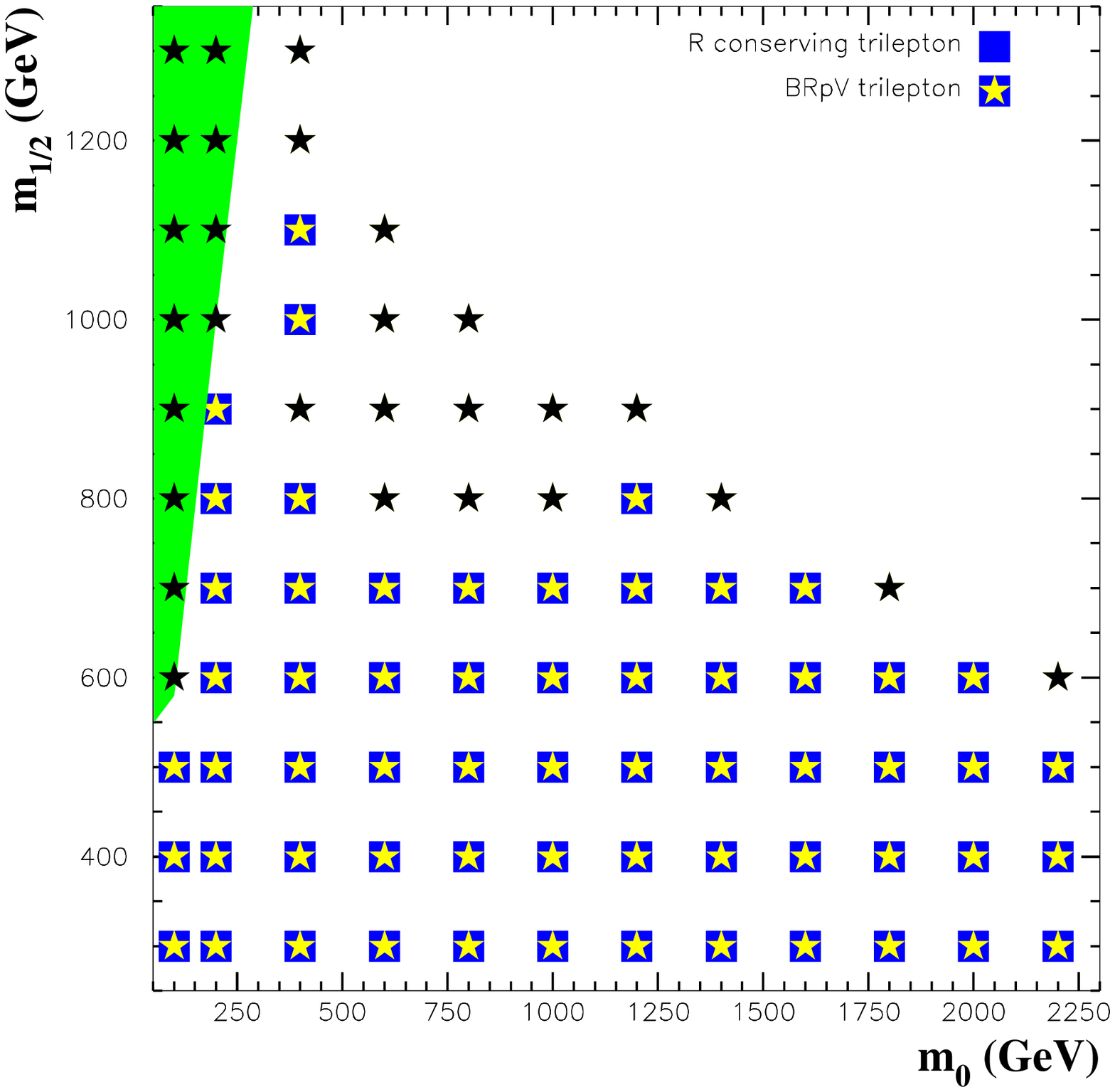}
 \includegraphics[width=7.5cm]{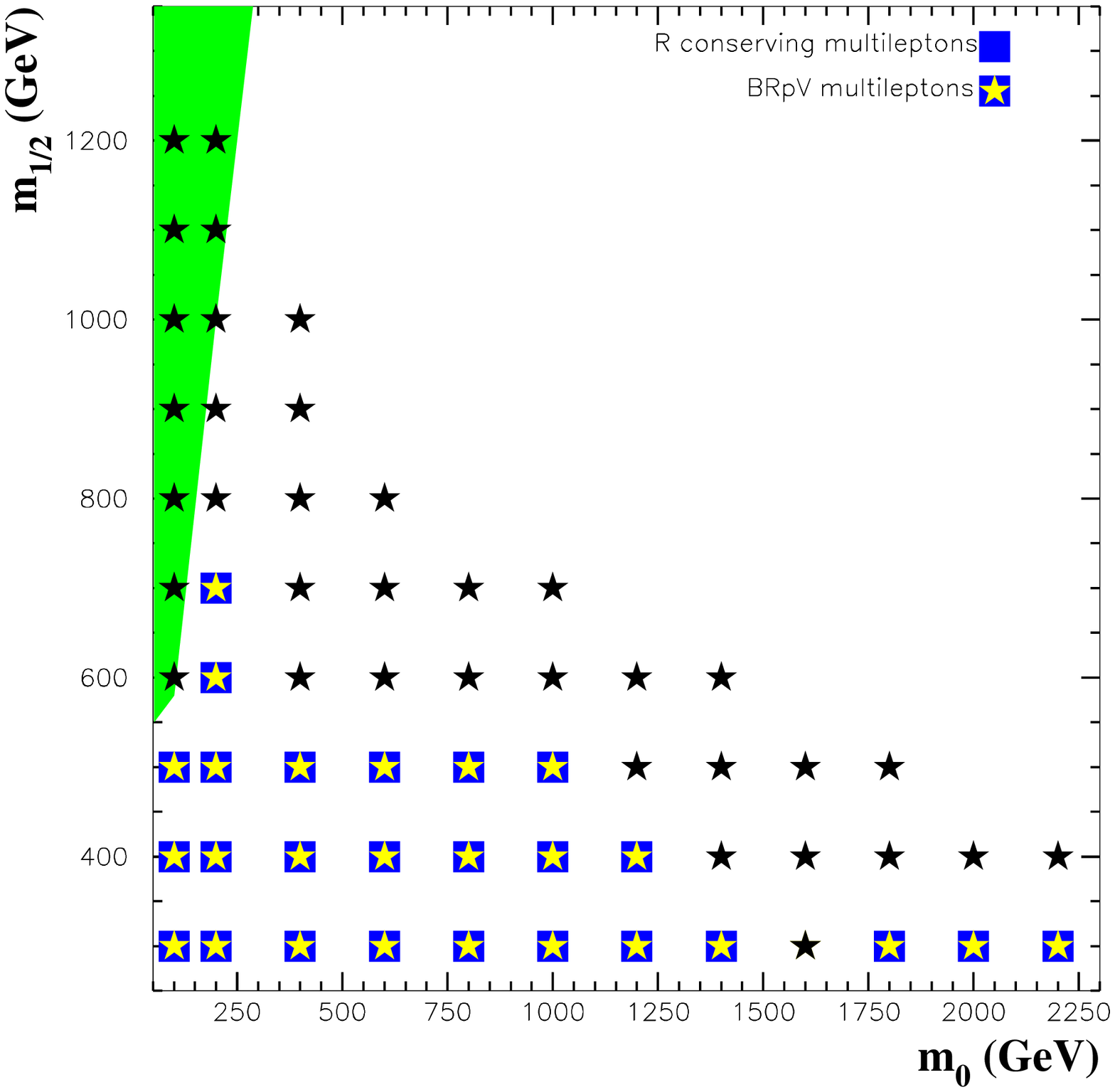}
  \end{center}
  \vspace*{-8mm}
  \caption{Same as Fig. \ref{fig:di:3lep10} but for an integrated
    luminosity of $100$ fb$^{-1}$.  }
\label{fig:di:3lep100}
\end{figure}

Figs.~\ref{fig:di:3lep10} and \ref{fig:di:3lep100} display the LHC
reach in the three-- (left panel) and multilepton (right--panel)
channels with/without R--parity conservation for integrated
luminosities of 10 and 100 fb$^{-1}$, respectively. The discovery
potential in these channels is rather limited in the R--parity
conserving scenario and it is enhanced when we add the BRpV
interactions. It is interesting to notice that the poor reach of the
$\mathbf{M\ell}$ topology at small $m_{1/2}$ ($\lesssim 200$ GeV) for
an integrated luminosity of 10 fb$^{-1}$ is due to the rather hard
values of $E_T^c$ used in our analysis.  Allowing $E_T^c = 100$ GeV
leads to a clear $\mathbf{M\ell}$ at small $m_{1/2}$.

For low integrated luminosities (10 fb$^{-1}$), the $\mathbf{3\ell}$
reach is smaller than that of the inclusive channel at small and
moderate $m_0$, however, it leads to a larger coverage at large $m_0$
in BRpV models.  This situation changes drastically at higher
luminosities with the trilepton signal becoming by far the main
channel for the search of supersymmetry with BRpV. The multilepton
channel also becomes very important, being the channel with the second
largest reach.

In brief, the decay of the lightest neutralino has an impact of
reducing the LHC reach for supersymmetry when we take into account
only the main all inclusive channel.  Notwithstanding, the new BRpV
interactions lead to a substantial increase in the final states
containing isolated charged leptons, specially for pairs of same sign
leptons and trileptons, compensating the losses in the inclusive
channel.  If supersymmetry is discovered in a set of the topologies
discussed above it must be a rather straightforward job to establish
the presence of R--parity violating interactions.  Now, we turn our
attention to another key characteristic feature of our BRpV--mSUGRA model,
namely, the emergence of displaced vertices associated to the LSP
decay.

\begin{figure}[th]
  \begin{center}
  \includegraphics[width=9.cm]{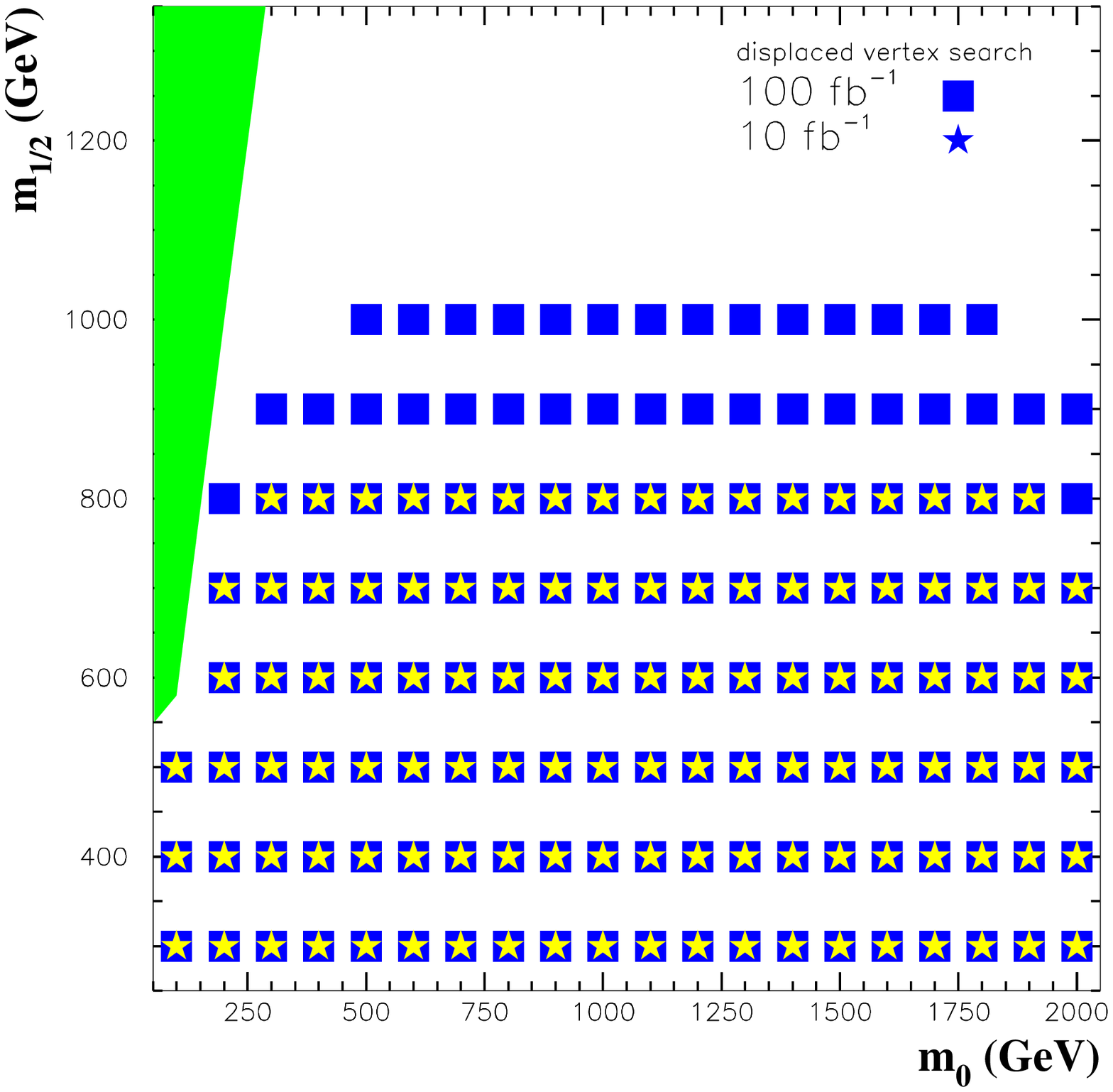}
  \end{center}
  \vspace*{-8mm}
  \caption{Discovery reach for displaced vertices channel in the
    $m_{0}\otimes m_{1/2}$ plane for $\tan\beta = 10$, $\mu > 0$,
    $A_0=-100$ GeV. The stars (squares) stand for points where there
    are more than 5 displaced vertex signal events for an integrated
    luminosity of 10 (100) fb$^{-1}$.  The marked grey (green) area on
    the left upper corner is the region where the stau is the LSP and
    the displaced vertex signal disappears. 
    Points already excluded by LEP and Tevatron searches are below the
    $m_{1/2}$ values depicted in this figure.  }
 \label{fig:ms}
\end{figure}

\subsection{Displaced Vertices Analysis}
\label{sec:displ-vert-analys-1}

Following the criteria previously used to define that a signal can be
observed in the canonical supersymmetry topologies, we have searched
for points of the parameter space presenting 5 or more identified
pairs of displaced vertices for integrated luminosities of 10 and 100
fb$^{-1}$, assuming that there is no SM background for this search.
Our results for the LHC discovery reach in the displaced vertex
channel are also shown in the $m_{0}\otimes m_{1/2}$ plane for
$\tan\beta = 10$, $\mu > 0$ and $A_0 = -100$ GeV.

In Figure~\ref{fig:ms} we present the displaced vertex reach.  As one
can see from this figure, the LHC will be able to look for the
displaced vertex signal up to $m_{1/2} \sim $800 (1000) GeV
($m_{\chi_{1}^{0}}\sim 340$ (430) GeV) for a large range of $m_0$ values and
an integrated luminosity of 10 (100) fb$^{-1}$. Notice that the reach in this
channel is rather independent of $m_0$ as expected from Fig.\ \ref{fig:ldec}.
Moreover, this signal for BRpV--mSUGRA 
disappears in the region where the stau is
the LSP due to its rather short lifetime. It is interesting to compare these
results with ones obtained in the canonical search channels.  The displaced
vertex signal provides the largest reach at moderate and large $m_0$, being
more important that the {\bf IN}, $\mathbf{3\ell}$ and $\mathbf{M\ell}$
topologies. Notwithstanding its good reach at small $m_0$, it is not the main
search channel in this range.  Certainly, the combination of the different
final state topologies will help us to pinpoint the properties of the model.
Furthermore, we should not forget that the discovery of high invariant mass
displaced vertices allow us to study directly the LSP such as its mass and
branching ratios. Moreover, we verified that displaced vertices possess a 
larger reach than the standard search channels not only at large $m_0$, but 
also at large $m_{1/2}$. We expect that for larger values of $\tan\beta$ the
reach in the displaced vertex analysis will be even better

Since the LHCb experiment has excellent vertex capabilities, it is natural to
conclude that the LHCb can also play an important part in the displaced vertex
search for supersymmetry, at least in the first stages of LHC operation, {\em
  i.e.}  when the accumulated luminosity is rather small~\cite{prepa}.

\section{Conclusions}

We have considered bilinear R--parity violating supergravity, the
simplest effective model for the supersymmetric origin of neutrino
masses, reproducing the current neutrino oscillation data.  We have
given a detailed analysis of the impact of R--parity violation on the
standard supersymmetry searches.  
We have examined how the LSP decays via bilinear R--parity violating
interactions degrade the average expected missing momentum in the
reactions and shown how this diminishes the reach in the main 
\emph{canonical} SUSY search channel that is the all inclusive topology.
However we have seen how this can be overcompensated by the coverage
obtained through the search for final states containing isolated
charged leptons, specially the trilepton signal. Moreover, we can also
search for displaced LSP decays which can further increase the LHC
reach for SUSY. We present the resulting reach in supergravity parameter
space for two reference luminosities of 10 \& 100 fb$^{-1}$.

In the first stages of LHC operation, when the accumulated luminosity
is rather small, we stress that a displaced-vertex analysis by the
LHCb collaboration can play an important role in probing bilinear
R--parity violating supergravity.
With enough luminosity to be collected in subsequent stages of
operation, the challenge will be to analyze the detailed nature of the
LSP decays as a test of the neutrino mixing angles measured in
oscillation experiments. We plan to turn to these issues elsewhere.


\section*{Acknowledgments}

We thank D. P. Roy for fruitful discussions.  Work supported by
Conselho Nacional de Desenvolvimento Cient\'{\i}fico e Tecnol\'ogico
(CNPq) and by Funda{c}\~ao de Amparo \`a Pesquisa do Estado de S\~ao
Paulo (FAPESP); by Spanish grants FPA2005-01269 and FPA2005-25348-E
(MEC), and ACOMP06/154 (Generalitat Valenciana), by European
Commission Contract MRTN-CT-2004-503369, by Colciencias in Colombia
under contract 1115-333-18740, and by the German Ministry of Education
and Research (BMBF) under contract 05HT6WWA.

\bibliographystyle{h-physrev}

\end{document}